\documentclass[prl,preprint,showpacs,groupedaddress]{revtex4-1}

\usepackage{graphicx}

\begin{document}


\title{Interatomic potential for the calculation of barrier distributions in amorphous oxides }


\author {J.P. Trinastic}
\author {R. Hamdan}
\author {Y. Wu}
\author {L. Zhang}
\author {Hai-Ping Cheng*}
 \affiliation {Department of Physics and Quantum Theory Project, University of Florida, Gainesville, Florida, 32611, USA}


\date{\today}

\begin{abstract}
Amorphous oxides are important for implants, optics, and gate insulators.  Understanding the effects of oxide doping is crucial to optimize performance. Here we report energy barrier distributions for amorphous tantala and doped oxides using a new set of computationally efficient, two-body potentials that reproduce the structural properties of the samples.  The distributions can be directly compared to experiment and used to calculate physical quantities such as internal friction.
\end{abstract}


\maketitle


Amorphous silica (SiO$_2$), tantala (Ta$_2$O$_5$), titania (TiO$_2$), and hafnia (HfO$_2$) are important oxides with applications as dielectrics in microelectronics, implants in medicine, and as mirror coatings for gravitational wave detection~\cite{Rowan200525,0264-9381-23-8-S10,0264-9381-21-5-004}, including the Laser Interferometer Gravitional Wave Observatory (LIGO) project~\cite{abramovici1992ligo}. In microelectronics, doping SiO$_2$ with oxides leads to high dielectric constants, equal performance and minimal leakage currents~\cite{Huff01012002,PhysRevLett.89.266101,0034-4885-69-2-R02,tomida:142902}. Tantala and Ta$_2$O$_5$-TiO$_2$ alloys are used in biomedical applications, such as artificial joints and stents, due to their superb biocompatibility~\cite{0957-4484-20-4-045102,JBM:JBM4,Bobyn1999347,Hou200398,Zhou200528}. Amorphous oxides used as mirror coatings for LIGO are the limiting factor for detection due to high mechanical loss (or internal friction)~\cite{0264-9381-19-5-304}. Titania-doped Ta$_2$O$_5$ and SiO$_2$ are the leading candidates to reduce loss, but more information about doping effects is needed~\cite{0264-9381-24-2-008,0264-9381-19-5-305,Gretarsson:07}.\newline
\indent Knowledge about how oxide doping affects physical characteristics such as thermal conductivity and internal friction is imperative to improve performance. In amorphous oxides, the lack of long-range order allows for the local rearrangements of atoms at temperatures below the glass-transition.  Thermally activated transitions between local energy minima associated with these rearrangements leads to unique acoustic and thermal properties.  These transitions have been characterized using an asymmetric double-well potential model to describe energy minima with asymmetry $\Delta$ separated by a barrier of height $V$~\cite{doi:10.1080/01418638108222343}.  It has been shown that properties such as specific heat~\cite{philips1972,nittke1995}, thermal conductivity~\cite{nittke1995}, and internal friction~\cite{doi:10.1080/01418638108222343} can be calculated once the distribution of energy barriers $g(V)$ is known for a given amorphous sample.  Therefore, the calculation of $g(V)$ for amorphous oxides is of fundamental importance to compare physical properties of oxides and doping levels.\newline
\indent Classical molecular dynamics (MD) is required to study these disordered systems. Previous computational work has calculated amorphous silica's barrier distribution using the Lennard Jones~\cite{heuer1997properties} and BKS potentials~\cite{reinisch2005moving}.  However, no molecular dynamics (MD) potentials exist for crystalline or amorphous Ta$_2$O$_5$ nor are there any potentials that combine all of the abovementioned oxides.  For the first time, we present barrier distributions for large scale, pure and doped amorphous systems using a novel set of  interatomic potentials for Ta$_2$O$_5$, TiO$_2$, and HfO$_2$ that can be used with previous potentials for  SiO$_2$~\cite{PhysRevLett.64.1955} and  zirconia (ZrO$_2$)~\cite{B902767J}. These distributions can be measured experimentally to compare with the present results.  The potentials have been designed with the goals of 1) transferability with each other and silica and 2) reproducing their bulk physical and elastic properties in crystalline and amorphous form.  After detailing the accuracy of the potentials, we report $g(V)$ for amorphous silica, tantala, and silica-doped hafnia and show doping-dependent changes.  The barriers are calculated using a bisection and ridge method~\cite{doliwa2003energy} that are discussed in detail elsewhere~\cite{hamdan_barriers}.\newline
\indent We employ the widely used BKS silica potential as a basis to develop the new potentials~\cite{PhysRevLett.64.1955}.  The BKS silica potential reproduces Si-O and O-O bond lengths as well as elastic properties of crystalline and amorphous silica.  Since the other oxides demonstrate similar O-O bond lengths (2.5-2.7 $\textup{\AA}$) in both crystalline and amorphous phases~\cite{Petkov199817,chen:114105,aleshina,wangcubichfo2}, this potential serves as a robust starting point to develop complementary two-body potentials for other materials using the same O-O potential parameters and charge number.  Although the BKS potential reproduces the main properties of silica, recent data on oxides such as  HfO$_2$ and TiO$_2$ suggest that the cation-anion bonding has covalent features~\cite{PhysRevB.65.174117,JACE:JACE1095}. In addition, experimental research has shown that sputtered and sol-gel-derived amorphous TiO$_2$ demonstrate predominantly six-coordinated and four-coordinated Ti atoms, respectively~\cite{Petkov199817}, suggesting different covalency based on amorphization procedure. Because the BKS potential does not model covalent bonding, we have fit the current potentials using an additional Morse term that controls the degree of covalency of the Ta-O, Ti-O, and Hf-O bonds.  Therefore, our final potential is a Morse-BKS (M-BKS) combined potential: ${\Phi}$$^{M-BKS}_{ij}$ = $q_i$$q_j$/$r_{ij}$ + $A_{ij}$$exp$(-$r_{ij}$/${\rho}_{ij}$) - $C_{ij}$/$r_{ij}^6$ + $D_{ij}$(1 - $exp$(-$a_{ij}$($r_{ij}$ - $r_e$))$^2$  (1), where ${\Phi}$$^{M-BKS}_{ij}$ is the total interaction energy between atoms i and j, $q_i$$q_j$/$r_{ij}$ is the Coulomb interaction, $A_{ij}$$exp$(-$r_{ij}$/${\rho}_{ij}$) represents the Pauli repulsion energy, $C_{ij}$/$r_{ij}^6$ represents the attractive van der Waals interaction, and $D_{ij}$(1 - $exp$(-$a_{ij}$($r_{ij}$ - $r_e$))$^2$ represents the covalent bond. The interaction energy is then a function of interatomic distance $r_{ij}$.  Fitting parameters $A_{ij}$, ${\rho}_{ij}$, $C_{ij}$, $D_{ij}$, $a_{ij}$, and $r_e$ are chosen to optimize lattice constants and elastic properties.  For silica, the charges $q_i$ and $q_j$ are 2.4 and -1.2 for Si and O, respectively, representing partial charge transfer between the atoms.  All cation charge numbers for the new potentials have been chosen to maintain charge neutrality with these values. Two-body potentials are advantageous due to their computational efficiency, and their simple construction makes it feasible to fit similar potentials for other materials to allow for doping studies. Previous potentials designed for these oxides are computationally expensive~\cite{herzbach:124711,PhysRevB.75.085311}, tailored for surface interactions~\cite{PhysRevB.75.085311}, or have not been fitted to mechanical properties~\cite{PhysRevLett.82.1708,Huff1999133,tangney:8898}. 

The optimal set of parameters for all interactions in the M-BKS potential are provided in Table I.  Potential parmaters have been fitted  to experimental and first-principles values of crystalline lattice constants and elastic moduli using the GULP software~\cite{doi:10.1080/0892702031000104887} (results compared to experimental and DFT results in Supplementary Materials).  Because most applications use their amorphous form, in the rest of the Letter we focus on amorphous samples created using the Lammps MD package (methods in Supplementary Materials)~\cite{Plimpton19951}.

The same M-BKS potential parameters have been used for the amorphous samples as in the crystalline polymorphs with one important addition related to titania. Because the differences in sputtered and sol-gel TiO$_2$ samples are related to coordination number, we propose that there is a difference in the strength of the covalent Ti-O bonds.  We represent a stronger covalency in the M-BKS potential by increasing the \textit{D$_{ij}$} parameter, deepening the potential well of the Morse interaction energy. Therefore, we fit a new potential parameter, Ti-O$_{Strong}$ reported in Table I, to represent a stronger covalent Ti-O bond for the six-coordinated amorphous titania samples ($D_{ij}$ = 0.5478 vs 0.3478). In the following, we compare samples of pure titania and titania-doped samples using either Ti-O interaction separately to examine how structural and elastic properties change.

Many possible doping combinations exist, however here we report structural and elastic properties of amorphous Ta$_2$O$_5$, TiO$_2$, and HfO$_2$, as well as TiO$_2$-doped Ta$_2$O$_5$, TiO$_2$-doped SiO$_2$, and SiO$_2$-doped HfO$_2$, which have experimental or first-principles data for comparison (summarized in Figures 1-3 and Table II). Figure 1(a) shows radial distribution functions of amorphous TiO$_2$ using either the Ti-O$_{Weak}$ or Ti-O$_{Strong}$ interaction. The weak-covalency Ti-O potential generates an amorphous structure with a Ti-O peak at 1.83 $\textup{\AA}$ and a Ti-O coordination number of 4.05 and Young's modulus (Y) of 73 GPa, closely matching previous experimental data for tetragonally coordinated titania produced using the sol-gel method (1.80-1.83 $\textup{\AA}$, 4-4.5 Ti-O coordination number, Y of 64 GPa)~\cite{varghese:193,Petkov199817,ottermann}.  On the other hand, the strong-covalency Ti-O potential produces an amorphous structure with a Ti-O peak at 1.96 $\textup{\AA}$, a Ti-O coordination number of 5.6, and Y of 170 GPa, reproducing recent data from sputtered amorphous titania (1.96 $\textup{\AA}$, 5.4 Ti-O coordination number, Y of ~160 GPa)~\cite{Petkov199817,ottermann,Zywitzki2004538}. This is the first MD potential to distinguish between the two types of amorphous titania based on amorphization procedure and also reproduce their elastic properties.

Little experimental or first-principles data exists about titania-doped silica, important for gate insulators~\cite{varghese:193}, however structural data regarding 12.5$\%$ titania-doped silica has been reported experimentally~\cite{Sandstrom1980201}.   Therefore, we dope amorphous silica with 12.5$\%$, 25$\%$, and 50$\%$ titania (Figure 1(b)-(d)).  Across all doping amounts, the silica network is undisturbed, as the Si-O bond length stays at 1.61 $\textup{\AA}$. Both titania potentials produce a similar structure at low titania content, however with increasing Ti the strong-covalency sample begins to form a harder, more eight-coordinated structure within the tetragonal silica network with a longer bond length.  On the other hand, the weak-covalency sample is already tetragonally coordinated and its bond length stays unchanged across all Ti doping levels.  Our results using the Ti-O$_{Weak}$ interaction matches previous experimental results of 12.5$\%$ titania doping of a largely tetragonal network~\cite{Sandstrom1980201}.

Silica-hafnia composites are also important materials for gate insulators due to hafnia's high dielectric constant.  A recent first-principles study has examined the structural properties of amorphous silica-doped hafnia~\cite{chen:114105}.  Therefore, we compare structural properties of pure hafnia as well as 25$\%$, 50$\%$, and 75$\%$ silica-doped hafnia to the first-principles results as another test of the transferability of the M-BKS potential.  As shown in Figure 2(a-d), the radial distribution functions for all four samples match very closely to the first-principles data, another confirmation of the robustness of the present potential.  Elastic properties for each sample are listed in Table II, however no previous data exists for comparison.

Finally, we examine the accuracy of our potential in reproducing amorphous tantala samples.  Experimental studies of amorphous pure and titania-doped tantala have calculated a reduced density function (RDF) from experimental diffraction data~\cite{bassiri2011probing,bassiri:031904,1742-6596-371-1-012058}, from which atomistic models are constructed using reverse Monte Carlo simulation.  From these models, we have calculated radial distribution functions to compare to our MD data in Figure 3 (RDF comparisons in Supplementary Materials).  For Ta$_2$O$_5$ (Figure 3(a)), the M-BKS potential reproduces the Ta-O peak at 1.94 $\textup{\AA}$ (Figure 3(a)) and a Ta-O coordination number of 5.85 (5.80-6.53 in experiment~\cite{banno:113507,bassiri:031904}).  The M-BKS potential generates a Ta-Ta peak at 3.74 $\textup{\AA}$, slightly longer than the experimental value of 3.67 $\textup{\AA}$.  The average O-O bond length in the sample is 2.77 $\textup{\AA}$, matching previous DFT results~\cite{Gu10072009}, and longer than the 2.6 $\textup{\AA}$ O-O bond length seen in silica.  This confirms the transferability of this two-body potential in distinguishing oxygen bond lengths between oxides while using the same O-O potential parameter.  The MD sample has a Young's modulus of 145 GPa, matching the experimental value of 140 GPa~\cite{martin2009comparison}.  This is the first MD potential to provide an accurate amorphous Ta$_2$O$_5$ structure and will be a valuable tool for future studies.

Results for TiO$_2$-doped Ta$_2$O$_5$ in Figure 3(b-e) show that the weak-covalency TiO$_2$ interaction leads to a titania peak around 1.85 $\textup{\AA}$, smaller than the 1.94 $\textup{\AA}$ peak seen experimentally.  On the other hand, using the strong-covalency TiO$_2$ interaction reproduces this 1.94 Ti-O $\textup{\AA}$ peak well, suggesting that titania is largely six-coordinated in the experimental data~\cite{evans_ta_ti}.  In both cases, the distribution functions do not significantly change between the 20.4 $\%$ and 53.8 $\%$ doping, and there is a good match between the M-BKS potential and experiment for the O-O and Ta-Ta radial distribution functions.  For the 53.8 $\%$ doping, the elastic moduli increase and decrease for the weak and strong Ti-O interactions, respectively. These results indicate that an amorphization method that creates four-coordinated Ti-O networks, such as sol-gel, could soften Ta$_2$O$_5$ samples.\newline
\indent  We next report the barrier distributions $g(V)$ for SiO$_2$, Ta$_2$O$_5$, and SiO$_2$-doped HfO$_2$ with an asymmery cutoff of $|\Delta|$ $\textless$ 52 meV (Figure 4). To characterize the barrier distribution, experiments traditionally report the distribution peak $h$, representing the dominant barrier affecting energy dissipation.  In addition, some experiments fit internal friction data assuming an exponential barrier distribution $g(V) = \frac{1}{V_0}exp(\frac{-V}{V_0})$ to calculate a  fitting parameter $V_0$ (see Supplementary Materials)~\cite{PhysRevLett.84.2718,travasso2007low,doi:10.1080/01418638108222343}. As seen from the logarithmic scales in Figure 4, our results indicate that the distributions demonstrate regimes with different exponential dependence, indicating that a single exponential fititng may not be accurate.  However, the data in Figure 4 can be used to calculate $V_0$ to compare with experiment as desired.  As shown in Figure 4(a), SiO$_2$ shows a peak at $h$ = 33.9 meV similar to the experimental finding of 44 meV~\cite{JACE:JACE125} and matching previous computational results~\cite{reinisch2005moving}. The barrier distribution for Ta$_2$O$_5$ (Figure 4(b)) demonstrates a peak at $h$ = 35.9 meV, matching experimental results of 28.6-42.0 meV similar to silica~\cite{martin2009comparison,0264-9381-27-22-225020,martin2008measurements}.

We next calculate the barrier distribution for  Si$_{0.75}$Hf$_{0.25}$O$_2$ and Si$_{0.50}$Hf$_{0.50}$O$_2$, reported in Figures 4(c) and 4(d), respectively, for which the distribution and peak are not known.  Compared to amorphous silica, the 75$\%$ silica-doped hafnia demonstrates a shift in the barrier distribution peak to 16.6 meV, however for 50$\%$ silica-doped hafnia the peak shifts back to 31.9 meV, similar to pure silica, indicating a nonmonotonic dependence on doping.  In SiO$_2$, the peak activation energy is believed to be due to Si-O bond angle shifts or elongations as well as SiO$_4$ tetrahedral rotations~\cite{JACE:JACE125}.  Due to a longer Hf-O bond length (2.04 vs 1.61 $\textup{\AA}$ in SiO$_2$), small amounts of HfO$_2$ may disrupt the tetrahedral network and lead to more accessible transitions. This is the first report of barrier distributions for tantala and doped oxide systems, and these result demonstrate that the present potentials can capture essential changes in the distributions with doping. Our future work will characterize the atomic reorientations associated with the distribution peaks and use the distributions to calculate physical properties.\newline
\indent In conclusion, we have presented the first computational study of barrier distributions for amorphous Ta$_2$O$_5$ and doped oxides using a new set of interatomic potentials that accurately reproduce structural and mechanical characteristics.  These distributions can be compared directly to experiment, and the results open the door for new computational studies to calculate physical properties of doped oxides to compare with and guide experimental findings. The distributions for SiO$_2$ and Ta$_2$O$_5$ replicate experimental results and distributions for SiO$_2$-doped HfO$_2$ show a nonmonotonic shift in the distribution peak with doping concentration. This work opens the way for MD studies of physical properties of doped oxide that were previously impossible due to a lack of transferable interatomic potentials.


%


\begin{figure}[h]
\includegraphics[scale=0.49]{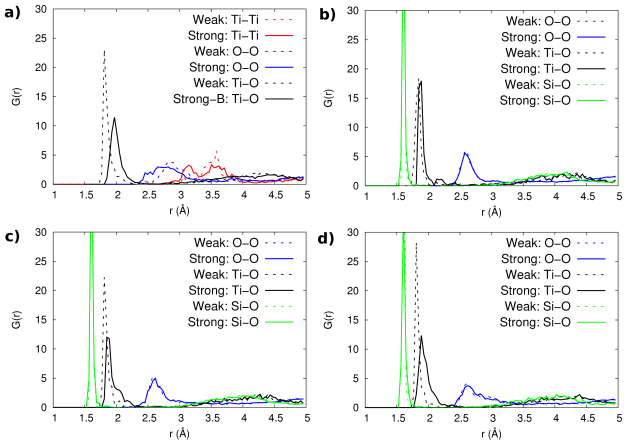}
\caption{(Color online) Radial distribution functions for amorphous titania and titania-doped silica comparing results using either the weak- or strong-covalent Ti-O interaction. a) TiO$_2$ b) 12.5$\%$ TiO$_2$. c) 25$\%$ TiO$_2$. d) 50$\%$ TiO$_2$}
\end{figure}

\begin{figure}[h]
\includegraphics[scale=0.49]{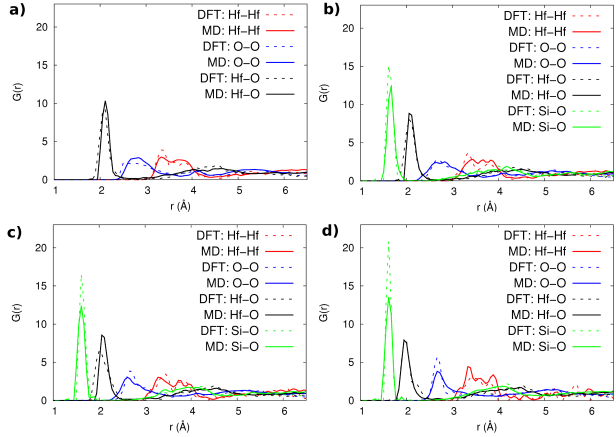}
\caption{(Color online) Radial distribution functions for amorphous hafnia and silica-doped hafnia using the M-BKS potential compared to previous DFT calculations~\cite{chen:114105}. a) HfO$_2$ b) 25$\%$ SiO$_2$. c) 50$\%$ SiO$_2$. d) 75$\%$ SiO$_2$.}
\end{figure}

\begin{figure}[h]
\includegraphics[scale=0.43]{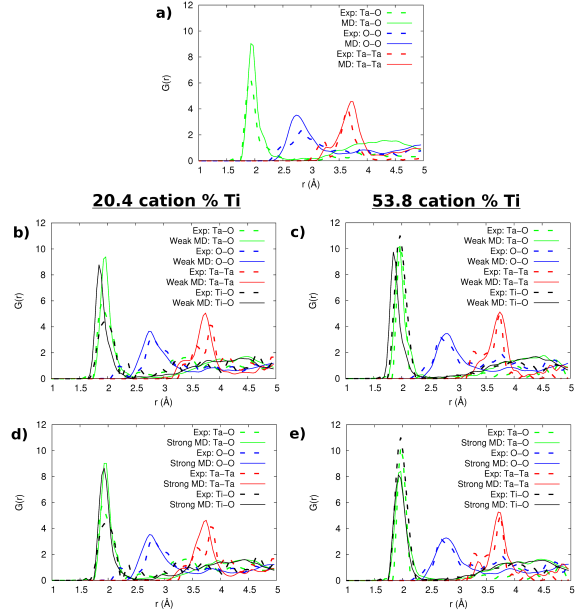}
\caption{(Color online) Radial distribution functions for amorphous pure and titania-doped tantala generated using the M-BKS potential compared to experiment~\cite{evans_ta_ti}. a) Ta$_2$O$_5$;  b) 20.4 cation $\%$ Ti using the Ti-O$_{Weak}$ interaction; c) 53.8 cation $\%$ Ti using the Ti-O$_{Weak}$ interaction; d) 20.4 cation $\%$ Ti using  the Ti-O$_{Strong}$ interaction; e) 53.8 cation $\%$ Ti using the Ti-O$_{Strong}$ interaction}
\end{figure}

\begin{figure}[h]
\includegraphics[scale=0.53]{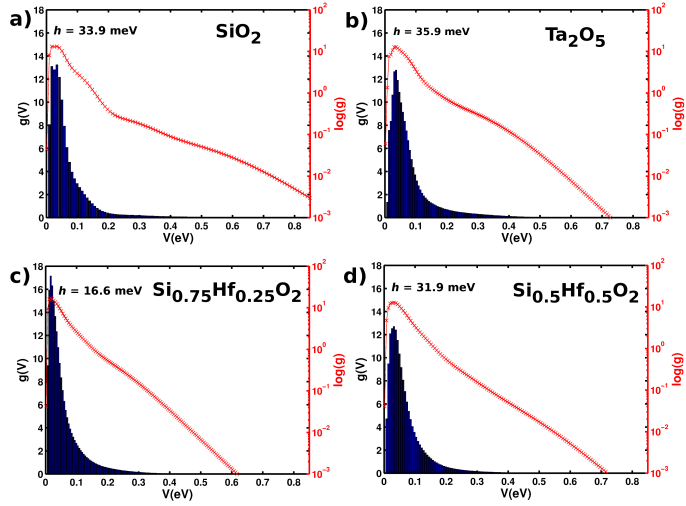}
\caption{(Color online) Normalized barrier distributions $g(V)$ (histogram, left y-scale) and $log(g)$ (red dotted line, right y-scale) as a function of barrier height $V$. The distribution peak value is labeled in each graph and discussed in the text. Fitting $g(V)$ to exponential curve $g(V) = \frac{1}{V_0}exp(\frac{-V}{V_0})$ provides $V_0$ fitting parameter to compare with experiment.  See text and Supplementary Materials for more discussion.  a) SiO$_2$;  b) Ta$_2$O$_5$; c) Si$_{0.75}$Hf$_{0.25}$O$_2$; d) Si$_{0.50}$Hf$_{0.50}$O$_2$}
\end{figure}


%

\begin{table}
\caption{Optimized parameters for the M-BKS potential.}
\resizebox{\columnwidth}{!}{%
\begin{tabular}{l*{7}{c}r}
Interaction & $A_{ij}$(eV) & ${\rho}_{ij}$($\textup{\AA}$) & $C_{ij}$(eV*$\AA^{6}$) & $D_{ij}$(eV) & $a_{ij}$($\AA^{-1}$) & $r_e$($\textup{\AA}$) & Charge (e) \\
\hline
O-O & 1388.77 & 0.3623 & 175.00 & 0.000 & 0.0000 & 0.0000 & O: -1.2 \\
Si-O & 18003.76 & 0.2052 & 133.54 & 0.000 & 0.0000 & 0.0000 & Si: 2.4 \\
Ta-O & 100067.01 & 0.1319 & 6.05 & 0.3789 & 1.6254 & 2.5445 & Ta: 3.0 \\
Hf-O & 12372.16 & 0.2286 & 81.36 & 0.3674 & 1.6230 & 2.0480 & Hf: 2.4 \\
Ti-O$_{Weak}$ & 5105.12 & 0.2253 & 20.000 & 0.3478& 1.9000 & 1.96000 & Ti: 2.4 \\
Ti-O$_{Strong}\footnote{This interaction is only relevant for differentiating amorphous samples.}$ & 5505.12 & 0.2253 & 20.000 & 0.5478 & 1.9000 & 1.9600 & Ti: 2.4 \\
\end{tabular}}
\end{table}

\begin{table}
\renewcommand{\tabcolsep}{0.30cm}
\small
\caption{Density, bulk modulus (B), shear modulus (G), and Young's modulus (Y) of amorphous samples (experimental or DFT values in parentheses when available)}
\resizebox{\columnwidth}{!}{%
\begin{tabular}{l*{15}{c}r}
\hline
  & Density (g/cm$^3$) & B (GPa) & G (GPa) & Y (GPa) & \\
\hline
Ta$_{2}$O$_{5}$-TiO$_2$ \\
\hspace{5mm} 0.0\% Ti & 8.10  & 118 & 56 & 145 \\
 & (5.92-8.00\footnote{~\cite{PhysRevB.83.144105}}) & & & (140$^a$) \\
\hspace{5mm} 20.4\% Ti & 7.04\footnote{Ti-O$_{weak}$ interaction}/7.20\footnote{Ti-O$_{strong}$ interaction} & 114$^b$/120$^c$ & 54$^b$/55$^c$ & 140$^b$/143$^c$ \\
\hspace{5mm} 53.8\% Ti & 5.60$^b$/6.01$^c$ & 94$^b$/132$^c$ & 47$^b$/63$^c$ & 122$^b$/163$^c$ \\
Si$_{1-x}$Ti$_x$O$_2$ \\
\hspace{5mm} x=0.125 &  2.58$^b$/2.68$^c$ & 42$^b$/56$^c$ & 27$^b$/29$^c$ & 67$^b$/74$^c$ \\
\hspace{5mm} x=0.25 &  2.61$^b$/2.87$^c$ & 36$^b$/62$^c$ & 24$^b$/32$^c$ & 59$^b$/82$^c$ \\
\hspace{5mm} x=0.50  &  2.68$^b$/3.19$^c$ & 32$^b$/73$^c$ & 18$^b$/35$^c$ & 46$^b$/91$^c$ \\
\hspace{5mm} x=1.00 &  2.92$^b$/3.75$^c$  & 54$^b$/140$^c$ & 29$^b$/66$^c$ & 73$^b$/170$^c$ \\
 & (2.9\footnote{~\cite{ottermann}}/3.8\footnote{~\cite{yinamorphoustio2}}) & & & (64$^d$/150-169\footnote{~\cite{Zywitzki2004538}}) \\
Si$_{1-x}$Hf$_x$O$_2$ \\
\hspace{5mm} x=0.25 & 4.08 & 64 & 38 & 94 \\
\hspace{5mm} x=0.50 & 5.64 & 73 & 45 & 112 \\
\hspace{5mm} x=1.00 & 8.67 & 145 & 59 & 155 \\
 & (8.8-9.6\footnote{~\cite{Triyoso01012004}})  & & \\
\hline
\end{tabular}}
\end{table}



\begin{acknowledgments}
\begin{center}\bf{Acknowledgments}\end{center}
*cheng@qtp.ufl.edu.  We acknolwedge support from NSF/PHY-1068138.   We also thank University of Florida High Performance Computing and NERSC for computing resources. 
\end{acknowledgments}
\bibliography{manuscript_incl_figures.bib}

\begin{thebibliography}{58}%
\makeatletter
\providecommand \@ifxundefined [1]{%
 \@ifx{#1\undefined}
}%
\providecommand \@ifnum [1]{%
 \ifnum #1\expandafter \@firstoftwo
 \else \expandafter \@secondoftwo
 \fi
}%
\providecommand \@ifx [1]{%
 \ifx #1\expandafter \@firstoftwo
 \else \expandafter \@secondoftwo
 \fi
}%
\providecommand \natexlab [1]{#1}%
\providecommand \enquote  [1]{``#1''}%
\providecommand \bibnamefont  [1]{#1}%
\providecommand \bibfnamefont [1]{#1}%
\providecommand \citenamefont [1]{#1}%
\providecommand \href@noop [0]{\@secondoftwo}%
\providecommand \href [0]{\begingroup \@sanitize@url \@href}%
\providecommand \@href[1]{\@@startlink{#1}\@@href}%
\providecommand \@@href[1]{\endgroup#1\@@endlink}%
\providecommand \@sanitize@url [0]{\catcode `\\12\catcode `\$12\catcode
  `\&12\catcode `\#12\catcode `\^12\catcode `\_12\catcode `\%12\relax}%
\providecommand \@@startlink[1]{}%
\providecommand \@@endlink[0]{}%
\providecommand \url  [0]{\begingroup\@sanitize@url \@url }%
\providecommand \@url [1]{\endgroup\@href {#1}{\urlprefix }}%
\providecommand \urlprefix  [0]{URL }%
\providecommand \Eprint [0]{\href }%
\providecommand \doibase [0]{http://dx.doi.org/}%
\providecommand \selectlanguage [0]{\@gobble}%
\providecommand \bibinfo  [0]{\@secondoftwo}%
\providecommand \bibfield  [0]{\@secondoftwo}%
\providecommand \translation [1]{[#1]}%
\providecommand \BibitemOpen [0]{}%
\providecommand \bibitemStop [0]{}%
\providecommand \bibitemNoStop [0]{.\EOS\space}%
\providecommand \EOS [0]{\spacefactor3000\relax}%
\providecommand \BibitemShut  [1]{\csname bibitem#1\endcsname}%
\let\auto@bib@innerbib\@empty
\bibitem [{\citenamefont {Rowan}\ \emph {et~al.}(2005)\citenamefont {Rowan},
  \citenamefont {Hough},\ and\ \citenamefont {Crooks}}]{Rowan200525}%
  \BibitemOpen
  \bibfield  {author} {\bibinfo {author} {\bibfnamefont {S.}~\bibnamefont
  {Rowan}}, \bibinfo {author} {\bibfnamefont {J.}~\bibnamefont {Hough}}, \ and\
  \bibinfo {author} {\bibfnamefont {D.}~\bibnamefont {Crooks}},\ }\href
  {\doibase 10.1016/j.physleta.2005.06.055} {\bibfield  {journal} {\bibinfo
  {journal} {Physics Letters A}\ }\textbf {\bibinfo {volume} {347}},\ \bibinfo
  {pages} {25 } (\bibinfo {year} {2005})}\BibitemShut {NoStop}%
\bibitem [{\citenamefont {Luck}\ \emph {et~al.}(2006)\citenamefont {Luck},
  \citenamefont {Hewitson}, \citenamefont {Ajith}, \citenamefont {Allen},
  \citenamefont {Aufmuth}, \citenamefont {Aulbert}, \citenamefont {Babak},
  \citenamefont {Balasubramanian}, \citenamefont {Barr}, \citenamefont
  {Berukoff}, \citenamefont {Bunkowski}, \citenamefont {Cagnoli}, \citenamefont
  {Cantley}, \citenamefont {Casey}, \citenamefont {Chelkowski}, \citenamefont
  {Chen}, \citenamefont {Churches}, \citenamefont {Cokelaer}, \citenamefont
  {Colacino}, \citenamefont {Crooks}, \citenamefont {Cutler}, \citenamefont
  {Danzmann}, \citenamefont {Dupuis}, \citenamefont {Elliffe}, \citenamefont
  {Fallnich}, \citenamefont {Franzen}, \citenamefont {Freise}, \citenamefont
  {Gholami}, \citenamefont {Goßler}, \citenamefont {Grant}, \citenamefont
  {Grote}, \citenamefont {Grunewald}, \citenamefont {Harms}, \citenamefont
  {Hage}, \citenamefont {Heinzel}, \citenamefont {Heng}, \citenamefont
  {Hepstonstall}, \citenamefont {Heurs}, \citenamefont {Hild}, \citenamefont
  {Hough}, \citenamefont {Itoh}, \citenamefont {Jones}, \citenamefont {Jones},
  \citenamefont {Huttner}, \citenamefont {Kötter}, \citenamefont {Krishnan},
  \citenamefont {Kwee}, \citenamefont {Luna}, \citenamefont {Machenschalk},
  \citenamefont {Malec}, \citenamefont {Mercer}, \citenamefont {Meier},
  \citenamefont {Messenger}, \citenamefont {Mohanty}, \citenamefont {Mossavi},
  \citenamefont {Mukherjee}, \citenamefont {Murray}, \citenamefont {Newton},
  \citenamefont {Papa}, \citenamefont {Perreur-Lloyd}, \citenamefont {Pitkin},
  \citenamefont {Plissi}, \citenamefont {Prix}, \citenamefont {Quetschke},
  \citenamefont {Re}, \citenamefont {Regimbau}, \citenamefont {Rehbein},
  \citenamefont {Reid}, \citenamefont {Ribichini}, \citenamefont {Robertson},
  \citenamefont {Robertson}, \citenamefont {Robinson}, \citenamefont {Romano},
  \citenamefont {Rowan}, \citenamefont {Rüdiger}, \citenamefont
  {Sathyaprakash}, \citenamefont {Schilling}, \citenamefont {Schnabel},
  \citenamefont {Schutz}, \citenamefont {Seifert}, \citenamefont {Sintes},
  \citenamefont {Smith}, \citenamefont {Sneddon}, \citenamefont {Strain},
  \citenamefont {Taylor}, \citenamefont {Taylor}, \citenamefont {Thüring},
  \citenamefont {Ungarelli}, \citenamefont {Vahlbruch}, \citenamefont
  {Vecchio}, \citenamefont {Veitch}, \citenamefont {Ward}, \citenamefont
  {Weiland}, \citenamefont {Welling}, \citenamefont {Wen}, \citenamefont
  {Williams}, \citenamefont {Willke}, \citenamefont {Winkler}, \citenamefont
  {Woan},\ and\ \citenamefont {Zhu}}]{0264-9381-23-8-S10}%
  \BibitemOpen
  \bibfield  {author} {\bibinfo {author} {\bibfnamefont {H.}~\bibnamefont
  {Luck}}, \bibinfo {author} {\bibfnamefont {M.}~\bibnamefont {Hewitson}},
  \bibinfo {author} {\bibfnamefont {P.}~\bibnamefont {Ajith}}, \bibinfo
  {author} {\bibfnamefont {B.}~\bibnamefont {Allen}}, \bibinfo {author}
  {\bibfnamefont {P.}~\bibnamefont {Aufmuth}}, \bibinfo {author} {\bibfnamefont
  {C.}~\bibnamefont {Aulbert}}, \bibinfo {author} {\bibfnamefont
  {S.}~\bibnamefont {Babak}}, \bibinfo {author} {\bibfnamefont
  {R.}~\bibnamefont {Balasubramanian}}, \bibinfo {author} {\bibfnamefont
  {B.~W.}\ \bibnamefont {Barr}}, \bibinfo {author} {\bibfnamefont
  {S.}~\bibnamefont {Berukoff}}, \bibinfo {author} {\bibfnamefont
  {A.}~\bibnamefont {Bunkowski}}, \bibinfo {author} {\bibfnamefont
  {G.}~\bibnamefont {Cagnoli}}, \bibinfo {author} {\bibfnamefont {C.~A.}\
  \bibnamefont {Cantley}}, \bibinfo {author} {\bibfnamefont {M.~M.}\
  \bibnamefont {Casey}}, \bibinfo {author} {\bibfnamefont {S.}~\bibnamefont
  {Chelkowski}}, \bibinfo {author} {\bibfnamefont {Y.}~\bibnamefont {Chen}},
  \bibinfo {author} {\bibfnamefont {D.}~\bibnamefont {Churches}}, \bibinfo
  {author} {\bibfnamefont {T.}~\bibnamefont {Cokelaer}}, \bibinfo {author}
  {\bibfnamefont {C.~N.}\ \bibnamefont {Colacino}}, \bibinfo {author}
  {\bibfnamefont {D.~R.~M.}\ \bibnamefont {Crooks}}, \bibinfo {author}
  {\bibfnamefont {C.}~\bibnamefont {Cutler}}, \bibinfo {author} {\bibfnamefont
  {K.}~\bibnamefont {Danzmann}}, \bibinfo {author} {\bibfnamefont {R.~J.}\
  \bibnamefont {Dupuis}}, \bibinfo {author} {\bibfnamefont {E.}~\bibnamefont
  {Elliffe}}, \bibinfo {author} {\bibfnamefont {C.}~\bibnamefont {Fallnich}},
  \bibinfo {author} {\bibfnamefont {A.}~\bibnamefont {Franzen}}, \bibinfo
  {author} {\bibfnamefont {A.}~\bibnamefont {Freise}}, \bibinfo {author}
  {\bibfnamefont {I.}~\bibnamefont {Gholami}}, \bibinfo {author} {\bibfnamefont
  {S.}~\bibnamefont {Goßler}}, \bibinfo {author} {\bibfnamefont
  {A.}~\bibnamefont {Grant}}, \bibinfo {author} {\bibfnamefont
  {H.}~\bibnamefont {Grote}}, \bibinfo {author} {\bibfnamefont
  {S.}~\bibnamefont {Grunewald}}, \bibinfo {author} {\bibfnamefont
  {J.}~\bibnamefont {Harms}}, \bibinfo {author} {\bibfnamefont
  {B.}~\bibnamefont {Hage}}, \bibinfo {author} {\bibfnamefont {G.}~\bibnamefont
  {Heinzel}}, \bibinfo {author} {\bibfnamefont {I.~S.}\ \bibnamefont {Heng}},
  \bibinfo {author} {\bibfnamefont {A.}~\bibnamefont {Hepstonstall}}, \bibinfo
  {author} {\bibfnamefont {M.}~\bibnamefont {Heurs}}, \bibinfo {author}
  {\bibfnamefont {S.}~\bibnamefont {Hild}}, \bibinfo {author} {\bibfnamefont
  {J.}~\bibnamefont {Hough}}, \bibinfo {author} {\bibfnamefont
  {Y.}~\bibnamefont {Itoh}}, \bibinfo {author} {\bibfnamefont {G.}~\bibnamefont
  {Jones}}, \bibinfo {author} {\bibfnamefont {R.}~\bibnamefont {Jones}},
  \bibinfo {author} {\bibfnamefont {S.~H.}\ \bibnamefont {Huttner}}, \bibinfo
  {author} {\bibfnamefont {K.}~\bibnamefont {Kötter}}, \bibinfo {author}
  {\bibfnamefont {B.}~\bibnamefont {Krishnan}}, \bibinfo {author}
  {\bibfnamefont {P.}~\bibnamefont {Kwee}}, \bibinfo {author} {\bibfnamefont
  {M.}~\bibnamefont {Luna}}, \bibinfo {author} {\bibfnamefont {B.}~\bibnamefont
  {Machenschalk}}, \bibinfo {author} {\bibfnamefont {M.}~\bibnamefont {Malec}},
  \bibinfo {author} {\bibfnamefont {R.~A.}\ \bibnamefont {Mercer}}, \bibinfo
  {author} {\bibfnamefont {T.}~\bibnamefont {Meier}}, \bibinfo {author}
  {\bibfnamefont {C.}~\bibnamefont {Messenger}}, \bibinfo {author}
  {\bibfnamefont {S.}~\bibnamefont {Mohanty}}, \bibinfo {author} {\bibfnamefont
  {K.}~\bibnamefont {Mossavi}}, \bibinfo {author} {\bibfnamefont
  {S.}~\bibnamefont {Mukherjee}}, \bibinfo {author} {\bibfnamefont
  {P.}~\bibnamefont {Murray}}, \bibinfo {author} {\bibfnamefont {G.~P.}\
  \bibnamefont {Newton}}, \bibinfo {author} {\bibfnamefont {M.~A.}\
  \bibnamefont {Papa}}, \bibinfo {author} {\bibfnamefont {M.}~\bibnamefont
  {Perreur-Lloyd}}, \bibinfo {author} {\bibfnamefont {M.}~\bibnamefont
  {Pitkin}}, \bibinfo {author} {\bibfnamefont {M.~V.}\ \bibnamefont {Plissi}},
  \bibinfo {author} {\bibfnamefont {R.}~\bibnamefont {Prix}}, \bibinfo {author}
  {\bibfnamefont {V.}~\bibnamefont {Quetschke}}, \bibinfo {author}
  {\bibfnamefont {V.}~\bibnamefont {Re}}, \bibinfo {author} {\bibfnamefont
  {T.}~\bibnamefont {Regimbau}}, \bibinfo {author} {\bibfnamefont
  {H.}~\bibnamefont {Rehbein}}, \bibinfo {author} {\bibfnamefont
  {S.}~\bibnamefont {Reid}}, \bibinfo {author} {\bibfnamefont {L.}~\bibnamefont
  {Ribichini}}, \bibinfo {author} {\bibfnamefont {D.~I.}\ \bibnamefont
  {Robertson}}, \bibinfo {author} {\bibfnamefont {N.~A.}\ \bibnamefont
  {Robertson}}, \bibinfo {author} {\bibfnamefont {C.}~\bibnamefont {Robinson}},
  \bibinfo {author} {\bibfnamefont {J.~D.}\ \bibnamefont {Romano}}, \bibinfo
  {author} {\bibfnamefont {S.}~\bibnamefont {Rowan}}, \bibinfo {author}
  {\bibfnamefont {A.}~\bibnamefont {Rüdiger}}, \bibinfo {author}
  {\bibfnamefont {B.~S.}\ \bibnamefont {Sathyaprakash}}, \bibinfo {author}
  {\bibfnamefont {R.}~\bibnamefont {Schilling}}, \bibinfo {author}
  {\bibfnamefont {R.}~\bibnamefont {Schnabel}}, \bibinfo {author}
  {\bibfnamefont {B.~F.}\ \bibnamefont {Schutz}}, \bibinfo {author}
  {\bibfnamefont {F.}~\bibnamefont {Seifert}}, \bibinfo {author} {\bibfnamefont
  {A.~M.}\ \bibnamefont {Sintes}}, \bibinfo {author} {\bibfnamefont {J.~R.}\
  \bibnamefont {Smith}}, \bibinfo {author} {\bibfnamefont {P.~H.}\ \bibnamefont
  {Sneddon}}, \bibinfo {author} {\bibfnamefont {K.~A.}\ \bibnamefont {Strain}},
  \bibinfo {author} {\bibfnamefont {I.}~\bibnamefont {Taylor}}, \bibinfo
  {author} {\bibfnamefont {R.}~\bibnamefont {Taylor}}, \bibinfo {author}
  {\bibfnamefont {A.}~\bibnamefont {Thüring}}, \bibinfo {author}
  {\bibfnamefont {C.}~\bibnamefont {Ungarelli}}, \bibinfo {author}
  {\bibfnamefont {H.}~\bibnamefont {Vahlbruch}}, \bibinfo {author}
  {\bibfnamefont {A.}~\bibnamefont {Vecchio}}, \bibinfo {author} {\bibfnamefont
  {J.}~\bibnamefont {Veitch}}, \bibinfo {author} {\bibfnamefont
  {H.}~\bibnamefont {Ward}}, \bibinfo {author} {\bibfnamefont {U.}~\bibnamefont
  {Weiland}}, \bibinfo {author} {\bibfnamefont {H.}~\bibnamefont {Welling}},
  \bibinfo {author} {\bibfnamefont {L.}~\bibnamefont {Wen}}, \bibinfo {author}
  {\bibfnamefont {P.}~\bibnamefont {Williams}}, \bibinfo {author}
  {\bibfnamefont {B.}~\bibnamefont {Willke}}, \bibinfo {author} {\bibfnamefont
  {W.}~\bibnamefont {Winkler}}, \bibinfo {author} {\bibfnamefont
  {G.}~\bibnamefont {Woan}}, \ and\ \bibinfo {author} {\bibfnamefont
  {R.}~\bibnamefont {Zhu}},\ }\href
  {http://stacks.iop.org/0264-9381/23/i=8/a=S10} {\bibfield  {journal}
  {\bibinfo  {journal} {Classical and Quantum Gravity}\ }\textbf {\bibinfo
  {volume} {23}},\ \bibinfo {pages} {S71} (\bibinfo {year} {2006})}\BibitemShut
  {NoStop}%
\bibitem [{\citenamefont {Takahashi}\ and\ \citenamefont {the
  TAMA~Collaboration}(2004)}]{0264-9381-21-5-004}%
  \BibitemOpen
  \bibfield  {author} {\bibinfo {author} {\bibfnamefont {R.}~\bibnamefont
  {Takahashi}}\ and\ \bibinfo {author} {\bibnamefont {the
  TAMA~Collaboration}},\ }\href {http://stacks.iop.org/0264-9381/21/i=5/a=004}
  {\bibfield  {journal} {\bibinfo  {journal} {Classical and Quantum Gravity}\
  }\textbf {\bibinfo {volume} {21}},\ \bibinfo {pages} {S403} (\bibinfo {year}
  {2004})}\BibitemShut {NoStop}%
\bibitem [{\citenamefont {Abramovici}\ \emph {et~al.}(1992)\citenamefont
  {Abramovici}, \citenamefont {Althouse}, \citenamefont {Drever}, \citenamefont
  {G{\"u}rsel}, \citenamefont {Kawamura}, \citenamefont {Raab}, \citenamefont
  {Shoemaker}, \citenamefont {Sievers}, \citenamefont {Spero}, \citenamefont
  {Thorne} \emph {et~al.}}]{abramovici1992ligo}%
  \BibitemOpen
  \bibfield  {author} {\bibinfo {author} {\bibfnamefont {A.}~\bibnamefont
  {Abramovici}}, \bibinfo {author} {\bibfnamefont {W.~E.}\ \bibnamefont
  {Althouse}}, \bibinfo {author} {\bibfnamefont {R.~W.}\ \bibnamefont
  {Drever}}, \bibinfo {author} {\bibfnamefont {Y.}~\bibnamefont {G{\"u}rsel}},
  \bibinfo {author} {\bibfnamefont {S.}~\bibnamefont {Kawamura}}, \bibinfo
  {author} {\bibfnamefont {F.~J.}\ \bibnamefont {Raab}}, \bibinfo {author}
  {\bibfnamefont {D.}~\bibnamefont {Shoemaker}}, \bibinfo {author}
  {\bibfnamefont {L.}~\bibnamefont {Sievers}}, \bibinfo {author} {\bibfnamefont
  {R.~E.}\ \bibnamefont {Spero}}, \bibinfo {author} {\bibfnamefont {K.~S.}\
  \bibnamefont {Thorne}},  \emph {et~al.},\ }\href@noop {} {\bibfield
  {journal} {\bibinfo  {journal} {Science}\ }\textbf {\bibinfo {volume}
  {256}},\ \bibinfo {pages} {325} (\bibinfo {year} {1992})}\BibitemShut
  {NoStop}%
\bibitem [{\citenamefont {Huff}(2002)}]{Huff01012002}%
  \BibitemOpen
  \bibfield  {author} {\bibinfo {author} {\bibfnamefont {H.~R.}\ \bibnamefont
  {Huff}},\ }\href {\doibase 10.1149/1.1471893} {\bibfield  {journal} {\bibinfo
   {journal} {Journal of The Electrochemical Society}\ }\textbf {\bibinfo
  {volume} {149}},\ \bibinfo {pages} {S35} (\bibinfo {year} {2002})},\ \Eprint
  {http://arxiv.org/abs/http://jes.ecsdl.org/content/149/5/S35.full.pdf+html}
  {http://jes.ecsdl.org/content/149/5/S35.full.pdf+html} \BibitemShut {NoStop}%
\bibitem [{\citenamefont {Fiorentini}\ and\ \citenamefont
  {Gulleri}(2002)}]{PhysRevLett.89.266101}%
  \BibitemOpen
  \bibfield  {author} {\bibinfo {author} {\bibfnamefont {V.}~\bibnamefont
  {Fiorentini}}\ and\ \bibinfo {author} {\bibfnamefont {G.}~\bibnamefont
  {Gulleri}},\ }\href {\doibase 10.1103/PhysRevLett.89.266101} {\bibfield
  {journal} {\bibinfo  {journal} {Phys. Rev. Lett.}\ }\textbf {\bibinfo
  {volume} {89}},\ \bibinfo {pages} {266101} (\bibinfo {year}
  {2002})}\BibitemShut {NoStop}%
\bibitem [{\citenamefont {Robertson}(2006)}]{0034-4885-69-2-R02}%
  \BibitemOpen
  \bibfield  {author} {\bibinfo {author} {\bibfnamefont {J.}~\bibnamefont
  {Robertson}},\ }\href {http://stacks.iop.org/0034-4885/69/i=2/a=R02}
  {\bibfield  {journal} {\bibinfo  {journal} {Reports on Progress in Physics}\
  }\textbf {\bibinfo {volume} {69}},\ \bibinfo {pages} {327} (\bibinfo {year}
  {2006})}\BibitemShut {NoStop}%
\bibitem [{\citenamefont {Tomida}\ \emph {et~al.}(2006)\citenamefont {Tomida},
  \citenamefont {Kita},\ and\ \citenamefont {Toriumi}}]{tomida:142902}%
  \BibitemOpen
  \bibfield  {author} {\bibinfo {author} {\bibfnamefont {K.}~\bibnamefont
  {Tomida}}, \bibinfo {author} {\bibfnamefont {K.}~\bibnamefont {Kita}}, \ and\
  \bibinfo {author} {\bibfnamefont {A.}~\bibnamefont {Toriumi}},\ }\href
  {\doibase 10.1063/1.2355471} {\bibfield  {journal} {\bibinfo  {journal}
  {Appl. Phys. Lett.}\ }\textbf {\bibinfo {volume} {89}},\ \bibinfo {eid}
  {142902} (\bibinfo {year} {2006})}\BibitemShut {NoStop}%
\bibitem [{\citenamefont {Ruckh}\ \emph {et~al.}(2009)\citenamefont {Ruckh},
  \citenamefont {Porter}, \citenamefont {Allam}, \citenamefont {Feng},
  \citenamefont {Grimes},\ and\ \citenamefont {Popat}}]{0957-4484-20-4-045102}%
  \BibitemOpen
  \bibfield  {author} {\bibinfo {author} {\bibfnamefont {T.}~\bibnamefont
  {Ruckh}}, \bibinfo {author} {\bibfnamefont {J.~R.}\ \bibnamefont {Porter}},
  \bibinfo {author} {\bibfnamefont {N.~K.}\ \bibnamefont {Allam}}, \bibinfo
  {author} {\bibfnamefont {X.}~\bibnamefont {Feng}}, \bibinfo {author}
  {\bibfnamefont {C.~A.}\ \bibnamefont {Grimes}}, \ and\ \bibinfo {author}
  {\bibfnamefont {K.~C.}\ \bibnamefont {Popat}},\ }\href
  {http://stacks.iop.org/0957-4484/20/i=4/a=045102} {\bibfield  {journal}
  {\bibinfo  {journal} {Nanotechnology}\ }\textbf {\bibinfo {volume} {20}},\
  \bibinfo {pages} {045102} (\bibinfo {year} {2009})}\BibitemShut {NoStop}%
\bibitem [{\citenamefont {Kato}\ \emph {et~al.}(2000)\citenamefont {Kato},
  \citenamefont {Nakamura}, \citenamefont {Nishiguchi}, \citenamefont
  {Matsusue}, \citenamefont {Kobayashi}, \citenamefont {Miyazaki},
  \citenamefont {Kim},\ and\ \citenamefont {Kokubo}}]{JBM:JBM4}%
  \BibitemOpen
  \bibfield  {author} {\bibinfo {author} {\bibfnamefont {H.}~\bibnamefont
  {Kato}}, \bibinfo {author} {\bibfnamefont {T.}~\bibnamefont {Nakamura}},
  \bibinfo {author} {\bibfnamefont {S.}~\bibnamefont {Nishiguchi}}, \bibinfo
  {author} {\bibfnamefont {Y.}~\bibnamefont {Matsusue}}, \bibinfo {author}
  {\bibfnamefont {M.}~\bibnamefont {Kobayashi}}, \bibinfo {author}
  {\bibfnamefont {T.}~\bibnamefont {Miyazaki}}, \bibinfo {author}
  {\bibfnamefont {H.}~\bibnamefont {Kim}}, \ and\ \bibinfo {author}
  {\bibfnamefont {T.}~\bibnamefont {Kokubo}},\ }\href {\doibase
  10.1002/(SICI)1097-4636(2000)53:1<28::AID-JBM4>3.0.CO;2-F} {\bibfield
  {journal} {\bibinfo  {journal} {Journal of Biomedical Materials Research}\
  }\textbf {\bibinfo {volume} {53}},\ \bibinfo {pages} {28} (\bibinfo {year}
  {2000})}\BibitemShut {NoStop}%
\bibitem [{\citenamefont {Bobyn}\ \emph {et~al.}(1999)\citenamefont {Bobyn},
  \citenamefont {Toh}, \citenamefont {Hacking}, \citenamefont {Tanzer},\ and\
  \citenamefont {Krygier}}]{Bobyn1999347}%
  \BibitemOpen
  \bibfield  {author} {\bibinfo {author} {\bibfnamefont {J.}~\bibnamefont
  {Bobyn}}, \bibinfo {author} {\bibfnamefont {K.-K.}\ \bibnamefont {Toh}},
  \bibinfo {author} {\bibfnamefont {S.}~\bibnamefont {Hacking}}, \bibinfo
  {author} {\bibfnamefont {M.}~\bibnamefont {Tanzer}}, \ and\ \bibinfo {author}
  {\bibfnamefont {J.~J.}\ \bibnamefont {Krygier}},\ }\href {\doibase
  10.1016/S0883-5403(99)90062-1} {\bibfield  {journal} {\bibinfo  {journal}
  {The Journal of Arthroplasty}\ }\textbf {\bibinfo {volume} {14}},\ \bibinfo
  {pages} {347 } (\bibinfo {year} {1999})}\BibitemShut {NoStop}%
\bibitem [{\citenamefont {Hou}\ \emph {et~al.}(2003)\citenamefont {Hou},
  \citenamefont {Zhuang}, \citenamefont {Zhang}, \citenamefont {Zhao},\ and\
  \citenamefont {Wu}}]{Hou200398}%
  \BibitemOpen
  \bibfield  {author} {\bibinfo {author} {\bibfnamefont {Y.-Q.}\ \bibnamefont
  {Hou}}, \bibinfo {author} {\bibfnamefont {D.-M.}\ \bibnamefont {Zhuang}},
  \bibinfo {author} {\bibfnamefont {G.}~\bibnamefont {Zhang}}, \bibinfo
  {author} {\bibfnamefont {M.}~\bibnamefont {Zhao}}, \ and\ \bibinfo {author}
  {\bibfnamefont {M.-S.}\ \bibnamefont {Wu}},\ }\href {\doibase
  10.1016/S0169-4332(03)00569-5} {\bibfield  {journal} {\bibinfo  {journal}
  {Applied Surface Science}\ }\textbf {\bibinfo {volume} {218}},\ \bibinfo
  {pages} {98 } (\bibinfo {year} {2003})}\BibitemShut {NoStop}%
\bibitem [{\citenamefont {Zhou}\ \emph {et~al.}(2005)\citenamefont {Zhou},
  \citenamefont {Niinomi}, \citenamefont {Akahori}, \citenamefont {Fukui},\
  and\ \citenamefont {Toda}}]{Zhou200528}%
  \BibitemOpen
  \bibfield  {author} {\bibinfo {author} {\bibfnamefont {Y.~L.}\ \bibnamefont
  {Zhou}}, \bibinfo {author} {\bibfnamefont {M.}~\bibnamefont {Niinomi}},
  \bibinfo {author} {\bibfnamefont {T.}~\bibnamefont {Akahori}}, \bibinfo
  {author} {\bibfnamefont {H.}~\bibnamefont {Fukui}}, \ and\ \bibinfo {author}
  {\bibfnamefont {H.}~\bibnamefont {Toda}},\ }\href {\doibase
  10.1016/j.msea.2005.03.032} {\bibfield  {journal} {\bibinfo  {journal}
  {Materials Science and Engineering: A}\ }\textbf {\bibinfo {volume} {398}},\
  \bibinfo {pages} {28 } (\bibinfo {year} {2005})}\BibitemShut {NoStop}%
\bibitem [{\citenamefont {Crooks}\ \emph {et~al.}(2002)\citenamefont {Crooks},
  \citenamefont {Sneddon}, \citenamefont {Cagnoli}, \citenamefont {Hough},
  \citenamefont {Rowan}, \citenamefont {Fejer}, \citenamefont {Gustafson},
  \citenamefont {Route}, \citenamefont {Nakagawa}, \citenamefont {Coyne},
  \citenamefont {Harry},\ and\ \citenamefont
  {Gretarsson}}]{0264-9381-19-5-304}%
  \BibitemOpen
  \bibfield  {author} {\bibinfo {author} {\bibfnamefont {D.~R.~M.}\
  \bibnamefont {Crooks}}, \bibinfo {author} {\bibfnamefont {P.}~\bibnamefont
  {Sneddon}}, \bibinfo {author} {\bibfnamefont {G.}~\bibnamefont {Cagnoli}},
  \bibinfo {author} {\bibfnamefont {J.}~\bibnamefont {Hough}}, \bibinfo
  {author} {\bibfnamefont {S.}~\bibnamefont {Rowan}}, \bibinfo {author}
  {\bibfnamefont {M.~M.}\ \bibnamefont {Fejer}}, \bibinfo {author}
  {\bibfnamefont {E.}~\bibnamefont {Gustafson}}, \bibinfo {author}
  {\bibfnamefont {R.}~\bibnamefont {Route}}, \bibinfo {author} {\bibfnamefont
  {N.}~\bibnamefont {Nakagawa}}, \bibinfo {author} {\bibfnamefont
  {D.}~\bibnamefont {Coyne}}, \bibinfo {author} {\bibfnamefont {G.~M.}\
  \bibnamefont {Harry}}, \ and\ \bibinfo {author} {\bibfnamefont {A.~M.}\
  \bibnamefont {Gretarsson}},\ }\href
  {http://stacks.iop.org/0264-9381/19/i=5/a=304} {\bibfield  {journal}
  {\bibinfo  {journal} {Classical and Quantum Gravity}\ }\textbf {\bibinfo
  {volume} {19}},\ \bibinfo {pages} {883} (\bibinfo {year} {2002})}\BibitemShut
  {NoStop}%
\bibitem [{\citenamefont {Harry}\ \emph {et~al.}(2007)\citenamefont {Harry},
  \citenamefont {Abernathy}, \citenamefont {Becerra-Toledo}, \citenamefont
  {Armandula}, \citenamefont {Black}, \citenamefont {Dooley}, \citenamefont
  {Eichenfield}, \citenamefont {Nwabugwu}, \citenamefont {Villar},
  \citenamefont {Crooks}, \citenamefont {Cagnoli}, \citenamefont {Hough},
  \citenamefont {How}, \citenamefont {MacLaren}, \citenamefont {Murray},
  \citenamefont {Reid}, \citenamefont {Rowan}, \citenamefont {Sneddon},
  \citenamefont {Fejer}, \citenamefont {Route}, \citenamefont {Penn},
  \citenamefont {Ganau}, \citenamefont {Mackowski}, \citenamefont {Michel},
  \citenamefont {Pinard},\ and\ \citenamefont
  {Remillieux}}]{0264-9381-24-2-008}%
  \BibitemOpen
  \bibfield  {author} {\bibinfo {author} {\bibfnamefont {G.~M.}\ \bibnamefont
  {Harry}}, \bibinfo {author} {\bibfnamefont {M.~R.}\ \bibnamefont
  {Abernathy}}, \bibinfo {author} {\bibfnamefont {A.~E.}\ \bibnamefont
  {Becerra-Toledo}}, \bibinfo {author} {\bibfnamefont {H.}~\bibnamefont
  {Armandula}}, \bibinfo {author} {\bibfnamefont {E.}~\bibnamefont {Black}},
  \bibinfo {author} {\bibfnamefont {K.}~\bibnamefont {Dooley}}, \bibinfo
  {author} {\bibfnamefont {M.}~\bibnamefont {Eichenfield}}, \bibinfo {author}
  {\bibfnamefont {C.}~\bibnamefont {Nwabugwu}}, \bibinfo {author}
  {\bibfnamefont {A.}~\bibnamefont {Villar}}, \bibinfo {author} {\bibfnamefont
  {D.~R.~M.}\ \bibnamefont {Crooks}}, \bibinfo {author} {\bibfnamefont
  {G.}~\bibnamefont {Cagnoli}}, \bibinfo {author} {\bibfnamefont
  {J.}~\bibnamefont {Hough}}, \bibinfo {author} {\bibfnamefont {C.~R.}\
  \bibnamefont {How}}, \bibinfo {author} {\bibfnamefont {I.}~\bibnamefont
  {MacLaren}}, \bibinfo {author} {\bibfnamefont {P.}~\bibnamefont {Murray}},
  \bibinfo {author} {\bibfnamefont {S.}~\bibnamefont {Reid}}, \bibinfo {author}
  {\bibfnamefont {S.}~\bibnamefont {Rowan}}, \bibinfo {author} {\bibfnamefont
  {P.~H.}\ \bibnamefont {Sneddon}}, \bibinfo {author} {\bibfnamefont {M.~M.}\
  \bibnamefont {Fejer}}, \bibinfo {author} {\bibfnamefont {R.}~\bibnamefont
  {Route}}, \bibinfo {author} {\bibfnamefont {S.~D.}\ \bibnamefont {Penn}},
  \bibinfo {author} {\bibfnamefont {P.}~\bibnamefont {Ganau}}, \bibinfo
  {author} {\bibfnamefont {J.-M.}\ \bibnamefont {Mackowski}}, \bibinfo {author}
  {\bibfnamefont {C.}~\bibnamefont {Michel}}, \bibinfo {author} {\bibfnamefont
  {L.}~\bibnamefont {Pinard}}, \ and\ \bibinfo {author} {\bibfnamefont
  {A.}~\bibnamefont {Remillieux}},\ }\href
  {http://stacks.iop.org/0264-9381/24/i=2/a=008} {\bibfield  {journal}
  {\bibinfo  {journal} {Classical and Quantum Gravity}\ }\textbf {\bibinfo
  {volume} {24}},\ \bibinfo {pages} {405} (\bibinfo {year} {2007})}\BibitemShut
  {NoStop}%
\bibitem [{\citenamefont {Harry}\ \emph {et~al.}(2002)\citenamefont {Harry},
  \citenamefont {Gretarsson}, \citenamefont {Saulson}, \citenamefont
  {Kittelberger}, \citenamefont {Penn}, \citenamefont {Startin}, \citenamefont
  {Rowan}, \citenamefont {Fejer}, \citenamefont {Crooks}, \citenamefont
  {Cagnoli}, \citenamefont {Hough},\ and\ \citenamefont
  {Nakagawa}}]{0264-9381-19-5-305}%
  \BibitemOpen
  \bibfield  {author} {\bibinfo {author} {\bibfnamefont {G.~M.}\ \bibnamefont
  {Harry}}, \bibinfo {author} {\bibfnamefont {A.~M.}\ \bibnamefont
  {Gretarsson}}, \bibinfo {author} {\bibfnamefont {P.~R.}\ \bibnamefont
  {Saulson}}, \bibinfo {author} {\bibfnamefont {S.~E.}\ \bibnamefont
  {Kittelberger}}, \bibinfo {author} {\bibfnamefont {S.~D.}\ \bibnamefont
  {Penn}}, \bibinfo {author} {\bibfnamefont {W.~J.}\ \bibnamefont {Startin}},
  \bibinfo {author} {\bibfnamefont {S.}~\bibnamefont {Rowan}}, \bibinfo
  {author} {\bibfnamefont {M.~M.}\ \bibnamefont {Fejer}}, \bibinfo {author}
  {\bibfnamefont {D.~R.~M.}\ \bibnamefont {Crooks}}, \bibinfo {author}
  {\bibfnamefont {G.}~\bibnamefont {Cagnoli}}, \bibinfo {author} {\bibfnamefont
  {J.}~\bibnamefont {Hough}}, \ and\ \bibinfo {author} {\bibfnamefont
  {N.}~\bibnamefont {Nakagawa}},\ }\href
  {http://stacks.iop.org/0264-9381/19/i=5/a=305} {\bibfield  {journal}
  {\bibinfo  {journal} {Classical and Quantum Gravity}\ }\textbf {\bibinfo
  {volume} {19}},\ \bibinfo {pages} {897} (\bibinfo {year} {2002})}\BibitemShut
  {NoStop}%
\bibitem [{\citenamefont {Gretarsson}\ \emph {et~al.}(2007)\citenamefont
  {Gretarsson}, \citenamefont {Harry}, \citenamefont {Ottaway}, \citenamefont
  {Agresti}, \citenamefont {Armandula}, \citenamefont {DeSalvo}, \citenamefont
  {Willems}, \citenamefont {Martin}, \citenamefont {Reid}, \citenamefont
  {Murray}, \citenamefont {Rowan}, \citenamefont {Hough}, \citenamefont
  {Fejer}, \citenamefont {Route}, \citenamefont {Penn}, \citenamefont {Pinto},
  \citenamefont {Galdi}, \citenamefont {Castaldi},\ and\ \citenamefont
  {Pierro}}]{Gretarsson:07}%
  \BibitemOpen
  \bibfield  {author} {\bibinfo {author} {\bibfnamefont {A.~M.}\ \bibnamefont
  {Gretarsson}}, \bibinfo {author} {\bibfnamefont {G.}~\bibnamefont {Harry}},
  \bibinfo {author} {\bibfnamefont {D.}~\bibnamefont {Ottaway}}, \bibinfo
  {author} {\bibfnamefont {J.}~\bibnamefont {Agresti}}, \bibinfo {author}
  {\bibfnamefont {H.}~\bibnamefont {Armandula}}, \bibinfo {author}
  {\bibfnamefont {R.}~\bibnamefont {DeSalvo}}, \bibinfo {author} {\bibfnamefont
  {P.}~\bibnamefont {Willems}}, \bibinfo {author} {\bibfnamefont
  {I.}~\bibnamefont {Martin}}, \bibinfo {author} {\bibfnamefont
  {S.}~\bibnamefont {Reid}}, \bibinfo {author} {\bibfnamefont {P.}~\bibnamefont
  {Murray}}, \bibinfo {author} {\bibfnamefont {S.}~\bibnamefont {Rowan}},
  \bibinfo {author} {\bibfnamefont {J.}~\bibnamefont {Hough}}, \bibinfo
  {author} {\bibfnamefont {M.}~\bibnamefont {Fejer}}, \bibinfo {author}
  {\bibfnamefont {R.}~\bibnamefont {Route}}, \bibinfo {author} {\bibfnamefont
  {S.}~\bibnamefont {Penn}}, \bibinfo {author} {\bibfnamefont {I.}~\bibnamefont
  {Pinto}}, \bibinfo {author} {\bibfnamefont {V.}~\bibnamefont {Galdi}},
  \bibinfo {author} {\bibfnamefont {G.}~\bibnamefont {Castaldi}}, \ and\
  \bibinfo {author} {\bibfnamefont {V.}~\bibnamefont {Pierro}},\ }in\ \href
  {http://www.opticsinfobase.org/abstract.cfm?URI=OIC-2007-ThC10} {\emph
  {\bibinfo {booktitle} {Optical Interference Coatings}}}\ (\bibinfo
  {publisher} {Optical Society of America},\ \bibinfo {year} {2007})\ p.\
  \bibinfo {pages} {ThC10}\BibitemShut {NoStop}%
\bibitem [{\citenamefont {Gilroy}\ and\ \citenamefont
  {Phillips}(1981)}]{doi:10.1080/01418638108222343}%
  \BibitemOpen
  \bibfield  {author} {\bibinfo {author} {\bibfnamefont {K.~S.}\ \bibnamefont
  {Gilroy}}\ and\ \bibinfo {author} {\bibfnamefont {W.~A.}\ \bibnamefont
  {Phillips}},\ }\href {\doibase 10.1080/01418638108222343} {\bibfield
  {journal} {\bibinfo  {journal} {Philosophical Magazine Part B}\ }\textbf
  {\bibinfo {volume} {43}},\ \bibinfo {pages} {735} (\bibinfo {year}
  {1981})}\BibitemShut {NoStop}%
\bibitem [{\citenamefont {Phillips}(1972)}]{philips1972}%
  \BibitemOpen
  \bibfield  {author} {\bibinfo {author} {\bibfnamefont {W.}~\bibnamefont
  {Phillips}},\ }\href {\doibase 10.1007/BF00660072} {\bibfield  {journal}
  {\bibinfo  {journal} {Journal of Low Temperature Physics}\ }\textbf {\bibinfo
  {volume} {7}},\ \bibinfo {pages} {351} (\bibinfo {year} {1972})}\BibitemShut
  {NoStop}%
\bibitem [{\citenamefont {Nittke}\ \emph {et~al.}(1995)\citenamefont {Nittke},
  \citenamefont {Scherl}, \citenamefont {Esquinazi}, \citenamefont {Lorenz},
  \citenamefont {Li},\ and\ \citenamefont {Pobell}}]{nittke1995}%
  \BibitemOpen
  \bibfield  {author} {\bibinfo {author} {\bibfnamefont {A.}~\bibnamefont
  {Nittke}}, \bibinfo {author} {\bibfnamefont {M.}~\bibnamefont {Scherl}},
  \bibinfo {author} {\bibfnamefont {P.}~\bibnamefont {Esquinazi}}, \bibinfo
  {author} {\bibfnamefont {W.}~\bibnamefont {Lorenz}}, \bibinfo {author}
  {\bibfnamefont {J.}~\bibnamefont {Li}}, \ and\ \bibinfo {author}
  {\bibfnamefont {F.}~\bibnamefont {Pobell}},\ }\href {\doibase
  10.1007/BF00752280} {\bibfield  {journal} {\bibinfo  {journal} {Journal of
  Low Temperature Physics}\ }\textbf {\bibinfo {volume} {98}},\ \bibinfo
  {pages} {517} (\bibinfo {year} {1995})}\BibitemShut {NoStop}%
\bibitem [{\citenamefont {Heuer}(1997)}]{heuer1997properties}%
  \BibitemOpen
  \bibfield  {author} {\bibinfo {author} {\bibfnamefont {A.}~\bibnamefont
  {Heuer}},\ }\href@noop {} {\bibfield  {journal} {\bibinfo  {journal} {Phys.
  Rev. Lett.}\ }\textbf {\bibinfo {volume} {78}},\ \bibinfo {pages} {4051}
  (\bibinfo {year} {1997})}\BibitemShut {NoStop}%
\bibitem [{\citenamefont {Reinisch}\ and\ \citenamefont
  {Heuer}(2005)}]{reinisch2005moving}%
  \BibitemOpen
  \bibfield  {author} {\bibinfo {author} {\bibfnamefont {J.}~\bibnamefont
  {Reinisch}}\ and\ \bibinfo {author} {\bibfnamefont {A.}~\bibnamefont
  {Heuer}},\ }\href@noop {} {\bibfield  {journal} {\bibinfo  {journal} {Phys.
  Rev. Lett.}\ }\textbf {\bibinfo {volume} {95}},\ \bibinfo {pages} {155502}
  (\bibinfo {year} {2005})}\BibitemShut {NoStop}%
\bibitem [{\citenamefont {van Beest}\ \emph {et~al.}(1990)\citenamefont {van
  Beest}, \citenamefont {Kramer},\ and\ \citenamefont {van
  Santen}}]{PhysRevLett.64.1955}%
  \BibitemOpen
  \bibfield  {author} {\bibinfo {author} {\bibfnamefont {B.~W.~H.}\
  \bibnamefont {van Beest}}, \bibinfo {author} {\bibfnamefont {G.~J.}\
  \bibnamefont {Kramer}}, \ and\ \bibinfo {author} {\bibfnamefont {R.~A.}\
  \bibnamefont {van Santen}},\ }\href {\doibase 10.1103/PhysRevLett.64.1955}
  {\bibfield  {journal} {\bibinfo  {journal} {Phys. Rev. Lett.}\ }\textbf
  {\bibinfo {volume} {64}},\ \bibinfo {pages} {1955} (\bibinfo {year}
  {1990})}\BibitemShut {NoStop}%
\bibitem [{\citenamefont {Yu}\ \emph {et~al.}(2009)\citenamefont {Yu},
  \citenamefont {Devanathan},\ and\ \citenamefont {Weber}}]{B902767J}%
  \BibitemOpen
  \bibfield  {author} {\bibinfo {author} {\bibfnamefont {J.}~\bibnamefont
  {Yu}}, \bibinfo {author} {\bibfnamefont {R.}~\bibnamefont {Devanathan}}, \
  and\ \bibinfo {author} {\bibfnamefont {W.~J.}\ \bibnamefont {Weber}},\ }\href
  {\doibase 10.1039/B902767J} {\bibfield  {journal} {\bibinfo  {journal} {J.
  Mater. Chem.}\ }\textbf {\bibinfo {volume} {19}},\ \bibinfo {pages} {3923}
  (\bibinfo {year} {2009})}\BibitemShut {NoStop}%
\bibitem [{\citenamefont {Doliwa}\ and\ \citenamefont
  {Heuer}(2003)}]{doliwa2003energy}%
  \BibitemOpen
  \bibfield  {author} {\bibinfo {author} {\bibfnamefont {B.}~\bibnamefont
  {Doliwa}}\ and\ \bibinfo {author} {\bibfnamefont {A.}~\bibnamefont {Heuer}},\
  }\href@noop {} {\bibfield  {journal} {\bibinfo  {journal} {Physical Review
  E}\ }\textbf {\bibinfo {volume} {67}},\ \bibinfo {pages} {031506} (\bibinfo
  {year} {2003})}\BibitemShut {NoStop}%
\bibitem [{\citenamefont {Hamdan}\ \emph {et~al.}()\citenamefont {Hamdan},
  \citenamefont {Trinastic},\ and\ \citenamefont {Cheng}}]{hamdan_barriers}%
  \BibitemOpen
  \bibfield  {author} {\bibinfo {author} {\bibfnamefont {R.}~\bibnamefont
  {Hamdan}}, \bibinfo {author} {\bibfnamefont {J.}~\bibnamefont {Trinastic}}, \
  and\ \bibinfo {author} {\bibfnamefont {H.~P.}\ \bibnamefont {Cheng}},\
  }\href@noop {} {\bibinfo  {journal} {in preparation}\ }\BibitemShut {NoStop}%
\bibitem [{\citenamefont {Petkov}\ \emph {et~al.}(1998)\citenamefont {Petkov},
  \citenamefont {HolzhÃ¼ter}, \citenamefont {TrÃ¶ge}, \citenamefont
  {Gerber},\ and\ \citenamefont {Himmel}}]{Petkov199817}%
  \BibitemOpen
\bibfield  {journal} {  }\bibfield  {author} {\bibinfo {author} {\bibfnamefont
  {V.}~\bibnamefont {Petkov}}, \bibinfo {author} {\bibfnamefont
  {G.}~\bibnamefont {HolzhÃ¼ter}}, \bibinfo {author} {\bibfnamefont
  {U.}~\bibnamefont {TrÃ¶ge}}, \bibinfo {author} {\bibfnamefont
  {T.}~\bibnamefont {Gerber}}, \ and\ \bibinfo {author} {\bibfnamefont
  {B.}~\bibnamefont {Himmel}},\ }\href {\doibase 10.1016/S0022-3093(98)00418-9}
  {\bibfield  {journal} {\bibinfo  {journal} {Journal of Non-Crystalline
  Solids}\ }\textbf {\bibinfo {volume} {231}},\ \bibinfo {pages} {17 }
  (\bibinfo {year} {1998})}\BibitemShut {NoStop}%
\bibitem [{\citenamefont {Chen}\ and\ \citenamefont {Kuo}(2011)}]{chen:114105}%
  \BibitemOpen
  \bibfield  {author} {\bibinfo {author} {\bibfnamefont {T.-J.}\ \bibnamefont
  {Chen}}\ and\ \bibinfo {author} {\bibfnamefont {C.-L.}\ \bibnamefont {Kuo}},\
  }\href {\doibase 10.1063/1.3664780} {\bibfield  {journal} {\bibinfo
  {journal} {Journal of Applied Physics}\ }\textbf {\bibinfo {volume} {110}},\
  \bibinfo {eid} {114105} (\bibinfo {year} {2011})}\BibitemShut {NoStop}%
\bibitem [{\citenamefont {Aleshina}\ and\ \citenamefont
  {Loginova}(2002)}]{aleshina}%
  \BibitemOpen
  \bibfield  {author} {\bibinfo {author} {\bibfnamefont {L.}~\bibnamefont
  {Aleshina}}\ and\ \bibinfo {author} {\bibfnamefont {S.}~\bibnamefont
  {Loginova}},\ }\href {\doibase 10.1134/1.1481927} {\bibfield  {journal}
  {\bibinfo  {journal} {Crystallography Reports}\ }\textbf {\bibinfo {volume}
  {47}},\ \bibinfo {pages} {415} (\bibinfo {year} {2002})}\BibitemShut
  {NoStop}%
\bibitem [{\citenamefont {Wang}\ \emph {et~al.}(1992)\citenamefont {Wang},
  \citenamefont {Li},\ and\ \citenamefont {Stevens}}]{wangcubichfo2}%
  \BibitemOpen
  \bibfield  {author} {\bibinfo {author} {\bibfnamefont {J.}~\bibnamefont
  {Wang}}, \bibinfo {author} {\bibfnamefont {H.}~\bibnamefont {Li}}, \ and\
  \bibinfo {author} {\bibfnamefont {R.}~\bibnamefont {Stevens}},\ }\href
  {\doibase 10.1007/BF00541601} {\bibfield  {journal} {\bibinfo  {journal}
  {Journal of Materials Science}\ }\textbf {\bibinfo {volume} {27}},\ \bibinfo
  {pages} {5397} (\bibinfo {year} {1992})}\BibitemShut {NoStop}%
\bibitem [{\citenamefont {Foster}\ \emph {et~al.}(2002)\citenamefont {Foster},
  \citenamefont {Lopez~Gejo}, \citenamefont {Shluger},\ and\ \citenamefont
  {Nieminen}}]{PhysRevB.65.174117}%
  \BibitemOpen
  \bibfield  {author} {\bibinfo {author} {\bibfnamefont {A.~S.}\ \bibnamefont
  {Foster}}, \bibinfo {author} {\bibfnamefont {F.}~\bibnamefont {Lopez~Gejo}},
  \bibinfo {author} {\bibfnamefont {A.~L.}\ \bibnamefont {Shluger}}, \ and\
  \bibinfo {author} {\bibfnamefont {R.~M.}\ \bibnamefont {Nieminen}},\ }\href
  {\doibase 10.1103/PhysRevB.65.174117} {\bibfield  {journal} {\bibinfo
  {journal} {Phys. Rev. B}\ }\textbf {\bibinfo {volume} {65}},\ \bibinfo
  {pages} {174117} (\bibinfo {year} {2002})}\BibitemShut {NoStop}%
\bibitem [{\citenamefont {Kim}\ \emph {et~al.}(1996)\citenamefont {Kim},
  \citenamefont {Enomoto}, \citenamefont {Nakagawa},\ and\ \citenamefont
  {Kawamura}}]{JACE:JACE1095}%
  \BibitemOpen
  \bibfield  {author} {\bibinfo {author} {\bibfnamefont {D.-W.}\ \bibnamefont
  {Kim}}, \bibinfo {author} {\bibfnamefont {N.}~\bibnamefont {Enomoto}},
  \bibinfo {author} {\bibfnamefont {Z.-e.}\ \bibnamefont {Nakagawa}}, \ and\
  \bibinfo {author} {\bibfnamefont {K.}~\bibnamefont {Kawamura}},\ }\href
  {\doibase 10.1111/j.1151-2916.1996.tb08553.x} {\bibfield  {journal} {\bibinfo
   {journal} {Journal of the American Ceramic Society}\ }\textbf {\bibinfo
  {volume} {79}},\ \bibinfo {pages} {1095} (\bibinfo {year}
  {1996})}\BibitemShut {NoStop}%
\bibitem [{\citenamefont {Herzbach}\ \emph {et~al.}(2005)\citenamefont
  {Herzbach}, \citenamefont {Binder},\ and\ \citenamefont
  {Muser}}]{herzbach:124711}%
  \BibitemOpen
  \bibfield  {author} {\bibinfo {author} {\bibfnamefont {D.}~\bibnamefont
  {Herzbach}}, \bibinfo {author} {\bibfnamefont {K.}~\bibnamefont {Binder}}, \
  and\ \bibinfo {author} {\bibfnamefont {M.~H.}\ \bibnamefont {Muser}},\ }\href
  {\doibase 10.1063/1.2038747} {\bibfield  {journal} {\bibinfo  {journal} {The
  Journal of Chemical Physics}\ }\textbf {\bibinfo {volume} {123}},\ \bibinfo
  {eid} {124711} (\bibinfo {year} {2005})}\BibitemShut {NoStop}%
\bibitem [{\citenamefont {Yu}\ \emph {et~al.}(2007)\citenamefont {Yu},
  \citenamefont {Sinnott},\ and\ \citenamefont
  {Phillpot}}]{PhysRevB.75.085311}%
  \BibitemOpen
  \bibfield  {author} {\bibinfo {author} {\bibfnamefont {J.}~\bibnamefont
  {Yu}}, \bibinfo {author} {\bibfnamefont {S.~B.}\ \bibnamefont {Sinnott}}, \
  and\ \bibinfo {author} {\bibfnamefont {S.~R.}\ \bibnamefont {Phillpot}},\
  }\href {\doibase 10.1103/PhysRevB.75.085311} {\bibfield  {journal} {\bibinfo
  {journal} {Phys. Rev. B}\ }\textbf {\bibinfo {volume} {75}},\ \bibinfo
  {pages} {085311} (\bibinfo {year} {2007})}\BibitemShut {NoStop}%
\bibitem [{\citenamefont {Demiralp}\ \emph {et~al.}(1999)\citenamefont
  {Demiralp}, \citenamefont {Cagin},\ and\ \citenamefont
  {Goddard}}]{PhysRevLett.82.1708}%
  \BibitemOpen
  \bibfield  {author} {\bibinfo {author} {\bibfnamefont {E.}~\bibnamefont
  {Demiralp}}, \bibinfo {author} {\bibfnamefont {T.}~\bibnamefont {Cagin}}, \
  and\ \bibinfo {author} {\bibfnamefont {W.~A.}\ \bibnamefont {Goddard}},\
  }\href {\doibase 10.1103/PhysRevLett.82.1708} {\bibfield  {journal} {\bibinfo
   {journal} {Phys. Rev. Lett.}\ }\textbf {\bibinfo {volume} {82}},\ \bibinfo
  {pages} {1708} (\bibinfo {year} {1999})}\BibitemShut {NoStop}%
\bibitem [{\citenamefont {Huff}\ \emph {et~al.}(1999)\citenamefont {Huff},
  \citenamefont {Demiralp}, \citenamefont {Cagin},\ and\ \citenamefont
  {III}}]{Huff1999133}%
  \BibitemOpen
  \bibfield  {author} {\bibinfo {author} {\bibfnamefont {N.~T.}\ \bibnamefont
  {Huff}}, \bibinfo {author} {\bibfnamefont {E.}~\bibnamefont {Demiralp}},
  \bibinfo {author} {\bibfnamefont {T.}~\bibnamefont {Cagin}}, \ and\ \bibinfo
  {author} {\bibfnamefont {W.~A.~G.}\ \bibnamefont {III}},\ }\href {\doibase
  10.1016/S0022-3093(99)00349-X} {\bibfield  {journal} {\bibinfo  {journal}
  {Journal of Non-Crystalline Solids}\ }\textbf {\bibinfo {volume} {253}},\
  \bibinfo {pages} {133 } (\bibinfo {year} {1999})}\BibitemShut {NoStop}%
\bibitem [{\citenamefont {Tangney}\ and\ \citenamefont
  {Scandolo}(2002)}]{tangney:8898}%
  \BibitemOpen
  \bibfield  {author} {\bibinfo {author} {\bibfnamefont {P.}~\bibnamefont
  {Tangney}}\ and\ \bibinfo {author} {\bibfnamefont {S.}~\bibnamefont
  {Scandolo}},\ }\href {\doibase 10.1063/1.1513312} {\bibfield  {journal}
  {\bibinfo  {journal} {The Journal of Chemical Physics}\ }\textbf {\bibinfo
  {volume} {117}},\ \bibinfo {pages} {8898} (\bibinfo {year}
  {2002})}\BibitemShut {NoStop}%
\bibitem [{\citenamefont {Gale}\ and\ \citenamefont
  {Rohl}(2003)}]{doi:10.1080/0892702031000104887}%
  \BibitemOpen
  \bibfield  {author} {\bibinfo {author} {\bibfnamefont {J.~D.}\ \bibnamefont
  {Gale}}\ and\ \bibinfo {author} {\bibfnamefont {A.~L.}\ \bibnamefont
  {Rohl}},\ }\href {\doibase 10.1080/0892702031000104887} {\bibfield  {journal}
  {\bibinfo  {journal} {Molecular Simulation}\ }\textbf {\bibinfo {volume}
  {29}},\ \bibinfo {pages} {291} (\bibinfo {year} {2003})}\BibitemShut
  {NoStop}%
\bibitem [{\citenamefont {Plimpton}(1995)}]{Plimpton19951}%
  \BibitemOpen
  \bibfield  {author} {\bibinfo {author} {\bibfnamefont {S.}~\bibnamefont
  {Plimpton}},\ }\href {\doibase 10.1006/jcph.1995.1039} {\bibfield  {journal}
  {\bibinfo  {journal} {Journal of Computational Physics}\ }\textbf {\bibinfo
  {volume} {117}},\ \bibinfo {pages} {1 } (\bibinfo {year} {1995})}\BibitemShut
  {NoStop}%
\bibitem [{\citenamefont {Varghese}\ \emph {et~al.}(2011)\citenamefont
  {Varghese}, \citenamefont {Joseph},\ and\ \citenamefont
  {Sebastian}}]{varghese:193}%
  \BibitemOpen
  \bibfield  {author} {\bibinfo {author} {\bibfnamefont {J.}~\bibnamefont
  {Varghese}}, \bibinfo {author} {\bibfnamefont {T.}~\bibnamefont {Joseph}}, \
  and\ \bibinfo {author} {\bibfnamefont {M.~T.}\ \bibnamefont {Sebastian}},\
  }\href {\doibase 10.1063/1.3644442} {\bibfield  {journal} {\bibinfo
  {journal} {AIP Conference Proceedings}\ }\textbf {\bibinfo {volume} {1372}},\
  \bibinfo {pages} {193} (\bibinfo {year} {2011})}\BibitemShut {NoStop}%
\bibitem [{\citenamefont {Ottermann}\ \emph {et~al.}()\citenamefont
  {Ottermann}, \citenamefont {Kuschnereit}, \citenamefont {Anderson},
  \citenamefont {Hess},\ and\ \citenamefont {Bange}}]{ottermann}%
  \BibitemOpen
  \bibfield  {author} {\bibinfo {author} {\bibfnamefont {C.~R.}\ \bibnamefont
  {Ottermann}}, \bibinfo {author} {\bibfnamefont {R.}~\bibnamefont
  {Kuschnereit}}, \bibinfo {author} {\bibfnamefont {O.}~\bibnamefont
  {Anderson}}, \bibinfo {author} {\bibfnamefont {P.}~\bibnamefont {Hess}}, \
  and\ \bibinfo {author} {\bibfnamefont {K.}~\bibnamefont {Bange}},\ }\href
  {\doibase 10.1557/PROC-436-251} {\bibfield  {journal} {\bibinfo  {journal}
  {MRS Proceedings}\ }\textbf {\bibinfo {volume} {436}},\
  10.1557/PROC-436-251}\BibitemShut {NoStop}%
\bibitem [{\citenamefont {Zywitzki}\ \emph {et~al.}(2004)\citenamefont
  {Zywitzki}, \citenamefont {Modes}, \citenamefont {Sahm}, \citenamefont
  {Frach}, \citenamefont {Goedicke},\ and\ \citenamefont
  {GlÃ¶ÃŸ}}]{Zywitzki2004538}%
  \BibitemOpen
  \bibfield  {author} {\bibinfo {author} {\bibfnamefont {O.}~\bibnamefont
  {Zywitzki}}, \bibinfo {author} {\bibfnamefont {T.}~\bibnamefont {Modes}},
  \bibinfo {author} {\bibfnamefont {H.}~\bibnamefont {Sahm}}, \bibinfo {author}
  {\bibfnamefont {P.}~\bibnamefont {Frach}}, \bibinfo {author} {\bibfnamefont
  {K.}~\bibnamefont {Goedicke}}, \ and\ \bibinfo {author} {\bibfnamefont
  {D.}~\bibnamefont {GlÃ¶ÃŸ}},\ }\href {\doibase
  10.1016/j.surfcoat.2003.10.115} {\bibfield  {journal} {\bibinfo  {journal}
  {Surface and Coatings Technology}\ }\textbf {\bibinfo {volume} {180-181}},\
  \bibinfo {pages} {538 } (\bibinfo {year} {2004})},\ \bibinfo {note}
  {proceedings of Symposium G on Protective Coatings and Thin Films-03, of the
  E-MRS 2003 Spring Conference}\BibitemShut {NoStop}%
\bibitem [{\citenamefont {Sandstrom}\ \emph {et~al.}(1980)\citenamefont
  {Sandstrom}, \citenamefont {Lytle}, \citenamefont {Wei}, \citenamefont
  {Greegor}, \citenamefont {Wong},\ and\ \citenamefont
  {Schultz}}]{Sandstrom1980201}%
  \BibitemOpen
  \bibfield  {author} {\bibinfo {author} {\bibfnamefont {D.~R.}\ \bibnamefont
  {Sandstrom}}, \bibinfo {author} {\bibfnamefont {F.~W.}\ \bibnamefont
  {Lytle}}, \bibinfo {author} {\bibfnamefont {P.}~\bibnamefont {Wei}}, \bibinfo
  {author} {\bibfnamefont {R.~B.}\ \bibnamefont {Greegor}}, \bibinfo {author}
  {\bibfnamefont {J.}~\bibnamefont {Wong}}, \ and\ \bibinfo {author}
  {\bibfnamefont {P.}~\bibnamefont {Schultz}},\ }\href {\doibase
  10.1016/0022-3093(80)90165-9} {\bibfield  {journal} {\bibinfo  {journal}
  {Journal of Non-Crystalline Solids}\ }\textbf {\bibinfo {volume} {41}},\
  \bibinfo {pages} {201 } (\bibinfo {year} {1980})}\BibitemShut {NoStop}%
\bibitem [{\citenamefont {Bassiri}\ \emph
  {et~al.}(2011{\natexlab{a}})\citenamefont {Bassiri}, \citenamefont
  {Borisenko}, \citenamefont {Cockayne}, \citenamefont {Hough}, \citenamefont
  {MacLaren},\ and\ \citenamefont {Rowan}}]{bassiri2011probing}%
  \BibitemOpen
  \bibfield  {author} {\bibinfo {author} {\bibfnamefont {R.}~\bibnamefont
  {Bassiri}}, \bibinfo {author} {\bibfnamefont {K.}~\bibnamefont {Borisenko}},
  \bibinfo {author} {\bibfnamefont {D.}~\bibnamefont {Cockayne}}, \bibinfo
  {author} {\bibfnamefont {J.}~\bibnamefont {Hough}}, \bibinfo {author}
  {\bibfnamefont {I.}~\bibnamefont {MacLaren}}, \ and\ \bibinfo {author}
  {\bibfnamefont {S.}~\bibnamefont {Rowan}},\ }\href@noop {} {\bibfield
  {journal} {\bibinfo  {journal} {Appl. Phys. Lett.}\ }\textbf {\bibinfo
  {volume} {98}},\ \bibinfo {pages} {031904} (\bibinfo {year}
  {2011}{\natexlab{a}})}\BibitemShut {NoStop}%
\bibitem [{\citenamefont {Bassiri}\ \emph
  {et~al.}(2011{\natexlab{b}})\citenamefont {Bassiri}, \citenamefont
  {Borisenko}, \citenamefont {Cockayne}, \citenamefont {Hough}, \citenamefont
  {MacLaren},\ and\ \citenamefont {Rowan}}]{bassiri:031904}%
  \BibitemOpen
  \bibfield  {author} {\bibinfo {author} {\bibfnamefont {R.}~\bibnamefont
  {Bassiri}}, \bibinfo {author} {\bibfnamefont {K.~B.}\ \bibnamefont
  {Borisenko}}, \bibinfo {author} {\bibfnamefont {D.~J.~H.}\ \bibnamefont
  {Cockayne}}, \bibinfo {author} {\bibfnamefont {J.}~\bibnamefont {Hough}},
  \bibinfo {author} {\bibfnamefont {I.}~\bibnamefont {MacLaren}}, \ and\
  \bibinfo {author} {\bibfnamefont {S.}~\bibnamefont {Rowan}},\ }\href
  {\doibase 10.1063/1.3535982} {\bibfield  {journal} {\bibinfo  {journal}
  {Appl. Phys. Lett.}\ }\textbf {\bibinfo {volume} {98}},\ \bibinfo {eid}
  {031904} (\bibinfo {year} {2011}{\natexlab{b}})}\BibitemShut {NoStop}%
\bibitem [{\citenamefont {Evans}\ \emph
  {et~al.}(2012{\natexlab{a}})\citenamefont {Evans}, \citenamefont {Bassiri},
  \citenamefont {Maclaren}, \citenamefont {Rowan}, \citenamefont {Martin},
  \citenamefont {Hough},\ and\ \citenamefont
  {Borisenko}}]{1742-6596-371-1-012058}%
  \BibitemOpen
  \bibfield  {author} {\bibinfo {author} {\bibfnamefont {K.}~\bibnamefont
  {Evans}}, \bibinfo {author} {\bibfnamefont {R.}~\bibnamefont {Bassiri}},
  \bibinfo {author} {\bibfnamefont {I.}~\bibnamefont {Maclaren}}, \bibinfo
  {author} {\bibfnamefont {S.}~\bibnamefont {Rowan}}, \bibinfo {author}
  {\bibfnamefont {I.}~\bibnamefont {Martin}}, \bibinfo {author} {\bibfnamefont
  {J.}~\bibnamefont {Hough}}, \ and\ \bibinfo {author} {\bibfnamefont {K.~B.}\
  \bibnamefont {Borisenko}},\ }\href
  {http://stacks.iop.org/1742-6596/371/i=1/a=012058} {\bibfield  {journal}
  {\bibinfo  {journal} {Journal of Physics: Conference Series}\ }\textbf
  {\bibinfo {volume} {371}},\ \bibinfo {pages} {012058} (\bibinfo {year}
  {2012}{\natexlab{a}})}\BibitemShut {NoStop}%
\bibitem [{\citenamefont {Banno}\ \emph {et~al.}(2010)\citenamefont {Banno},
  \citenamefont {Sakamoto}, \citenamefont {Iguchi}, \citenamefont {Matsumoto},
  \citenamefont {Imai}, \citenamefont {Ichihashi}, \citenamefont {Fujieda},
  \citenamefont {Tanaka}, \citenamefont {Watanabe}, \citenamefont {Yamaguchi},
  \citenamefont {Hasegawa},\ and\ \citenamefont {Aono}}]{banno:113507}%
  \BibitemOpen
  \bibfield  {author} {\bibinfo {author} {\bibfnamefont {N.}~\bibnamefont
  {Banno}}, \bibinfo {author} {\bibfnamefont {T.}~\bibnamefont {Sakamoto}},
  \bibinfo {author} {\bibfnamefont {N.}~\bibnamefont {Iguchi}}, \bibinfo
  {author} {\bibfnamefont {M.}~\bibnamefont {Matsumoto}}, \bibinfo {author}
  {\bibfnamefont {H.}~\bibnamefont {Imai}}, \bibinfo {author} {\bibfnamefont
  {T.}~\bibnamefont {Ichihashi}}, \bibinfo {author} {\bibfnamefont
  {S.}~\bibnamefont {Fujieda}}, \bibinfo {author} {\bibfnamefont
  {K.}~\bibnamefont {Tanaka}}, \bibinfo {author} {\bibfnamefont
  {S.}~\bibnamefont {Watanabe}}, \bibinfo {author} {\bibfnamefont
  {S.}~\bibnamefont {Yamaguchi}}, \bibinfo {author} {\bibfnamefont
  {T.}~\bibnamefont {Hasegawa}}, \ and\ \bibinfo {author} {\bibfnamefont
  {M.}~\bibnamefont {Aono}},\ }\href {\doibase 10.1063/1.3488830} {\bibfield
  {journal} {\bibinfo  {journal} {Appl. Phys. Lett.}\ }\textbf {\bibinfo
  {volume} {97}},\ \bibinfo {eid} {113507} (\bibinfo {year}
  {2010})}\BibitemShut {NoStop}%
\bibitem [{\citenamefont {Gu}\ and\ \citenamefont
  {Watanabe}(2009)}]{Gu10072009}%
  \BibitemOpen
  \bibfield  {author} {\bibinfo {author} {\bibfnamefont {T.}~\bibnamefont
  {Gu}}\ and\ \bibinfo {author} {\bibfnamefont {S.}~\bibnamefont {Watanabe}},\
  }\href {http://ma.ecsdl.org/content/MA2009-02/2/167.abstract} {\bibfield
  {journal} {\bibinfo  {journal} {Meeting Abstracts}\ }\textbf {\bibinfo
  {volume} {MA2009-02}},\ \bibinfo {pages} {167} (\bibinfo {year} {2009})},\
  \Eprint
  {http://arxiv.org/abs/http://ma.ecsdl.org/content/MA2009-02/2/167.full.pdf+html}
  {http://ma.ecsdl.org/content/MA2009-02/2/167.full.pdf+html} \BibitemShut
  {NoStop}%
\bibitem [{\citenamefont {Martin}\ \emph {et~al.}(2009)\citenamefont {Martin},
  \citenamefont {Chalkley}, \citenamefont {Nawrodt}, \citenamefont {Armandula},
  \citenamefont {Bassiri}, \citenamefont {Comtet}, \citenamefont {Fejer},
  \citenamefont {Gretarsson}, \citenamefont {Harry}, \citenamefont {Heinert}
  \emph {et~al.}}]{martin2009comparison}%
  \BibitemOpen
  \bibfield  {author} {\bibinfo {author} {\bibfnamefont {I.}~\bibnamefont
  {Martin}}, \bibinfo {author} {\bibfnamefont {E.}~\bibnamefont {Chalkley}},
  \bibinfo {author} {\bibfnamefont {R.}~\bibnamefont {Nawrodt}}, \bibinfo
  {author} {\bibfnamefont {H.}~\bibnamefont {Armandula}}, \bibinfo {author}
  {\bibfnamefont {R.}~\bibnamefont {Bassiri}}, \bibinfo {author} {\bibfnamefont
  {C.}~\bibnamefont {Comtet}}, \bibinfo {author} {\bibfnamefont
  {M.}~\bibnamefont {Fejer}}, \bibinfo {author} {\bibfnamefont
  {A.}~\bibnamefont {Gretarsson}}, \bibinfo {author} {\bibfnamefont
  {G.}~\bibnamefont {Harry}}, \bibinfo {author} {\bibfnamefont
  {D.}~\bibnamefont {Heinert}},  \emph {et~al.},\ }\href@noop {} {\bibfield
  {journal} {\bibinfo  {journal} {Classical and Quantum Gravity}\ }\textbf
  {\bibinfo {volume} {26}},\ \bibinfo {pages} {155012} (\bibinfo {year}
  {2009})}\BibitemShut {NoStop}%
\bibitem [{\citenamefont {Evans}\ \emph
  {et~al.}(2012{\natexlab{b}})\citenamefont {Evans}, \citenamefont {Bassiri},
  \citenamefont {Maclaren}, \citenamefont {Rowan}, \citenamefont {Martin},
  \citenamefont {Hough},\ and\ \citenamefont {Borisenko}}]{evans_ta_ti}%
  \BibitemOpen
  \bibfield  {author} {\bibinfo {author} {\bibfnamefont {K.}~\bibnamefont
  {Evans}}, \bibinfo {author} {\bibfnamefont {R.}~\bibnamefont {Bassiri}},
  \bibinfo {author} {\bibfnamefont {I.}~\bibnamefont {Maclaren}}, \bibinfo
  {author} {\bibfnamefont {S.}~\bibnamefont {Rowan}}, \bibinfo {author}
  {\bibfnamefont {I.}~\bibnamefont {Martin}}, \bibinfo {author} {\bibfnamefont
  {J.}~\bibnamefont {Hough}}, \ and\ \bibinfo {author} {\bibfnamefont {K.~B.}\
  \bibnamefont {Borisenko}},\ }\href
  {http://stacks.iop.org/1742-6596/371/i=1/a=012058} {\bibfield  {journal}
  {\bibinfo  {journal} {Journal of Physics: Conference Series}\ }\textbf
  {\bibinfo {volume} {371}},\ \bibinfo {pages} {012058} (\bibinfo {year}
  {2012}{\natexlab{b}})}\BibitemShut {NoStop}%
\bibitem [{\citenamefont {Wiedersich}\ \emph {et~al.}(2000)\citenamefont
  {Wiedersich}, \citenamefont {Adichtchev},\ and\ \citenamefont
  {R\"ossler}}]{PhysRevLett.84.2718}%
  \BibitemOpen
  \bibfield  {author} {\bibinfo {author} {\bibfnamefont {J.}~\bibnamefont
  {Wiedersich}}, \bibinfo {author} {\bibfnamefont {S.~V.}\ \bibnamefont
  {Adichtchev}}, \ and\ \bibinfo {author} {\bibfnamefont {E.}~\bibnamefont
  {R\"ossler}},\ }\href {\doibase 10.1103/PhysRevLett.84.2718} {\bibfield
  {journal} {\bibinfo  {journal} {Phys. Rev. Lett.}\ }\textbf {\bibinfo
  {volume} {84}},\ \bibinfo {pages} {2718} (\bibinfo {year}
  {2000})}\BibitemShut {NoStop}%
\bibitem [{\citenamefont {Travasso}\ \emph {et~al.}(2007)\citenamefont
  {Travasso}, \citenamefont {Amico}, \citenamefont {Bosi}, \citenamefont
  {Cottone}, \citenamefont {Dari}, \citenamefont {Gammaitoni}, \citenamefont
  {Vocca},\ and\ \citenamefont {Marchesoni}}]{travasso2007low}%
  \BibitemOpen
  \bibfield  {author} {\bibinfo {author} {\bibfnamefont {F.}~\bibnamefont
  {Travasso}}, \bibinfo {author} {\bibfnamefont {P.}~\bibnamefont {Amico}},
  \bibinfo {author} {\bibfnamefont {L.}~\bibnamefont {Bosi}}, \bibinfo {author}
  {\bibfnamefont {F.}~\bibnamefont {Cottone}}, \bibinfo {author} {\bibfnamefont
  {A.}~\bibnamefont {Dari}}, \bibinfo {author} {\bibfnamefont {L.}~\bibnamefont
  {Gammaitoni}}, \bibinfo {author} {\bibfnamefont {H.}~\bibnamefont {Vocca}}, \
  and\ \bibinfo {author} {\bibfnamefont {F.}~\bibnamefont {Marchesoni}},\
  }\href@noop {} {\bibfield  {journal} {\bibinfo  {journal} {EPL (Europhysics
  Letters)}\ }\textbf {\bibinfo {volume} {80}},\ \bibinfo {pages} {50008}
  (\bibinfo {year} {2007})}\BibitemShut {NoStop}%
\bibitem [{\citenamefont {Anderson}\ and\ \citenamefont
  {Bommel}(1955)}]{JACE:JACE125}%
  \BibitemOpen
  \bibfield  {author} {\bibinfo {author} {\bibfnamefont {O.~L.}\ \bibnamefont
  {Anderson}}\ and\ \bibinfo {author} {\bibfnamefont {H.~E.}\ \bibnamefont
  {Bommel}},\ }\href {\doibase 10.1111/j.1151-2916.1955.tb14914.x} {\bibfield
  {journal} {\bibinfo  {journal} {Journal of the American Ceramic Society}\
  }\textbf {\bibinfo {volume} {38}},\ \bibinfo {pages} {125} (\bibinfo {year}
  {1955})}\BibitemShut {NoStop}%
\bibitem [{\citenamefont {Martin}\ \emph {et~al.}(2010)\citenamefont {Martin},
  \citenamefont {Bassiri}, \citenamefont {Nawrodt}, \citenamefont {Fejer},
  \citenamefont {Gretarsson}, \citenamefont {Gustafson}, \citenamefont {Harry},
  \citenamefont {Hough}, \citenamefont {MacLaren}, \citenamefont {Penn},
  \citenamefont {Reid}, \citenamefont {Route}, \citenamefont {Rowan},
  \citenamefont {Schwarz}, \citenamefont {Seidel}, \citenamefont {Scott},\ and\
  \citenamefont {Woodcraft}}]{0264-9381-27-22-225020}%
  \BibitemOpen
  \bibfield  {author} {\bibinfo {author} {\bibfnamefont {I.~W.}\ \bibnamefont
  {Martin}}, \bibinfo {author} {\bibfnamefont {R.}~\bibnamefont {Bassiri}},
  \bibinfo {author} {\bibfnamefont {R.}~\bibnamefont {Nawrodt}}, \bibinfo
  {author} {\bibfnamefont {M.~M.}\ \bibnamefont {Fejer}}, \bibinfo {author}
  {\bibfnamefont {A.}~\bibnamefont {Gretarsson}}, \bibinfo {author}
  {\bibfnamefont {E.}~\bibnamefont {Gustafson}}, \bibinfo {author}
  {\bibfnamefont {G.}~\bibnamefont {Harry}}, \bibinfo {author} {\bibfnamefont
  {J.}~\bibnamefont {Hough}}, \bibinfo {author} {\bibfnamefont
  {I.}~\bibnamefont {MacLaren}}, \bibinfo {author} {\bibfnamefont
  {S.}~\bibnamefont {Penn}}, \bibinfo {author} {\bibfnamefont {S.}~\bibnamefont
  {Reid}}, \bibinfo {author} {\bibfnamefont {R.}~\bibnamefont {Route}},
  \bibinfo {author} {\bibfnamefont {S.}~\bibnamefont {Rowan}}, \bibinfo
  {author} {\bibfnamefont {C.}~\bibnamefont {Schwarz}}, \bibinfo {author}
  {\bibfnamefont {P.}~\bibnamefont {Seidel}}, \bibinfo {author} {\bibfnamefont
  {J.}~\bibnamefont {Scott}}, \ and\ \bibinfo {author} {\bibfnamefont {A.~L.}\
  \bibnamefont {Woodcraft}},\ }\href
  {http://stacks.iop.org/0264-9381/27/i=22/a=225020} {\bibfield  {journal}
  {\bibinfo  {journal} {Classical and Quantum Gravity}\ }\textbf {\bibinfo
  {volume} {27}},\ \bibinfo {pages} {225020} (\bibinfo {year}
  {2010})}\BibitemShut {NoStop}%
\bibitem [{\citenamefont {Martin}\ \emph {et~al.}(2008)\citenamefont {Martin},
  \citenamefont {Armandula}, \citenamefont {Comtet}, \citenamefont {Fejer},
  \citenamefont {Gretarsson}, \citenamefont {Harry}, \citenamefont {Hough},
  \citenamefont {Mackowski}, \citenamefont {MacLaren}, \citenamefont {Michel}
  \emph {et~al.}}]{martin2008measurements}%
  \BibitemOpen
  \bibfield  {author} {\bibinfo {author} {\bibfnamefont {I.}~\bibnamefont
  {Martin}}, \bibinfo {author} {\bibfnamefont {H.}~\bibnamefont {Armandula}},
  \bibinfo {author} {\bibfnamefont {C.}~\bibnamefont {Comtet}}, \bibinfo
  {author} {\bibfnamefont {M.}~\bibnamefont {Fejer}}, \bibinfo {author}
  {\bibfnamefont {A.}~\bibnamefont {Gretarsson}}, \bibinfo {author}
  {\bibfnamefont {G.}~\bibnamefont {Harry}}, \bibinfo {author} {\bibfnamefont
  {J.}~\bibnamefont {Hough}}, \bibinfo {author} {\bibfnamefont {J.~M.}\
  \bibnamefont {Mackowski}}, \bibinfo {author} {\bibfnamefont {I.}~\bibnamefont
  {MacLaren}}, \bibinfo {author} {\bibfnamefont {C.}~\bibnamefont {Michel}},
  \emph {et~al.},\ }\href@noop {} {\bibfield  {journal} {\bibinfo  {journal}
  {Classical and Quantum gravity}\ }\textbf {\bibinfo {volume} {25}},\ \bibinfo
  {pages} {055005} (\bibinfo {year} {2008})}\BibitemShut {NoStop}%
\bibitem [{\citenamefont {Wu}\ \emph {et~al.}(2011)\citenamefont {Wu},
  \citenamefont {Li},\ and\ \citenamefont {Cheng}}]{PhysRevB.83.144105}%
  \BibitemOpen
  \bibfield  {author} {\bibinfo {author} {\bibfnamefont {Y.-N.}\ \bibnamefont
  {Wu}}, \bibinfo {author} {\bibfnamefont {L.}~\bibnamefont {Li}}, \ and\
  \bibinfo {author} {\bibfnamefont {H.-P.}\ \bibnamefont {Cheng}},\ }\href
  {\doibase 10.1103/PhysRevB.83.144105} {\bibfield  {journal} {\bibinfo
  {journal} {Phys. Rev. B}\ }\textbf {\bibinfo {volume} {83}},\ \bibinfo
  {pages} {144105} (\bibinfo {year} {2011})}\BibitemShut {NoStop}%
\bibitem [{\citenamefont {Yin}\ \emph {et~al.}(2001)\citenamefont {Yin},
  \citenamefont {Wada}, \citenamefont {Kitamura}, \citenamefont {Kambe},
  \citenamefont {Murasawa}, \citenamefont {Mori}, \citenamefont {Sakata},\ and\
  \citenamefont {Yanagida}}]{yinamorphoustio2}%
  \BibitemOpen
  \bibfield  {author} {\bibinfo {author} {\bibfnamefont {H.}~\bibnamefont
  {Yin}}, \bibinfo {author} {\bibfnamefont {Y.}~\bibnamefont {Wada}}, \bibinfo
  {author} {\bibfnamefont {T.}~\bibnamefont {Kitamura}}, \bibinfo {author}
  {\bibfnamefont {S.}~\bibnamefont {Kambe}}, \bibinfo {author} {\bibfnamefont
  {S.}~\bibnamefont {Murasawa}}, \bibinfo {author} {\bibfnamefont
  {H.}~\bibnamefont {Mori}}, \bibinfo {author} {\bibfnamefont {T.}~\bibnamefont
  {Sakata}}, \ and\ \bibinfo {author} {\bibfnamefont {S.}~\bibnamefont
  {Yanagida}},\ }\href {\doibase 10.1039/B008974P} {\bibfield  {journal}
  {\bibinfo  {journal} {J. Mater. Chem.}\ }\textbf {\bibinfo {volume} {11}},\
  \bibinfo {pages} {1694} (\bibinfo {year} {2001})}\BibitemShut {NoStop}%
\bibitem [{\citenamefont {Triyoso}\ \emph {et~al.}(2004)\citenamefont
  {Triyoso}, \citenamefont {Liu}, \citenamefont {Roan}, \citenamefont {Ramon},
  \citenamefont {Edwards}, \citenamefont {Gregory}, \citenamefont {Werho},
  \citenamefont {Kulik}, \citenamefont {Tam}, \citenamefont {Irwin},
  \citenamefont {Wang}, \citenamefont {La}, \citenamefont {Hobbs},
  \citenamefont {Garcia}, \citenamefont {Baker}, \citenamefont {White},\ and\
  \citenamefont {Tobin}}]{Triyoso01012004}%
  \BibitemOpen
  \bibfield  {author} {\bibinfo {author} {\bibfnamefont {D.}~\bibnamefont
  {Triyoso}}, \bibinfo {author} {\bibfnamefont {R.}~\bibnamefont {Liu}},
  \bibinfo {author} {\bibfnamefont {D.}~\bibnamefont {Roan}}, \bibinfo {author}
  {\bibfnamefont {M.}~\bibnamefont {Ramon}}, \bibinfo {author} {\bibfnamefont
  {N.~V.}\ \bibnamefont {Edwards}}, \bibinfo {author} {\bibfnamefont
  {R.}~\bibnamefont {Gregory}}, \bibinfo {author} {\bibfnamefont
  {D.}~\bibnamefont {Werho}}, \bibinfo {author} {\bibfnamefont
  {J.}~\bibnamefont {Kulik}}, \bibinfo {author} {\bibfnamefont
  {G.}~\bibnamefont {Tam}}, \bibinfo {author} {\bibfnamefont {E.}~\bibnamefont
  {Irwin}}, \bibinfo {author} {\bibfnamefont {X.-D.}\ \bibnamefont {Wang}},
  \bibinfo {author} {\bibfnamefont {L.~B.}\ \bibnamefont {La}}, \bibinfo
  {author} {\bibfnamefont {C.}~\bibnamefont {Hobbs}}, \bibinfo {author}
  {\bibfnamefont {R.}~\bibnamefont {Garcia}}, \bibinfo {author} {\bibfnamefont
  {J.}~\bibnamefont {Baker}}, \bibinfo {author} {\bibfnamefont {B.~E.}\
  \bibnamefont {White}}, \ and\ \bibinfo {author} {\bibfnamefont
  {P.}~\bibnamefont {Tobin}},\ }\href {\doibase 10.1149/1.1784821} {\bibfield
  {journal} {\bibinfo  {journal} {Journal of The Electrochemical Society}\
  }\textbf {\bibinfo {volume} {151}},\ \bibinfo {pages} {F220} (\bibinfo {year}
  {2004})},\ \Eprint
  {http://arxiv.org/abs/http://jes.ecsdl.org/content/151/10/F220.full.pdf+html}
  {http://jes.ecsdl.org/content/151/10/F220.full.pdf+html} \BibitemShut
  {NoStop}%
\end{thebibliography}%

\end{document}